\documentclass[a4paper,12pt]{article}

\pdfoutput=1

\usepackage{multirow}
\usepackage{amsmath}
\usepackage{tgtermes}
\usepackage{xcolor}
\usepackage{amsfonts}
\usepackage{amssymb}
\usepackage{amstext}
\usepackage{import,breqn}
\usepackage{graphicx}
\usepackage{caption,subcaption}
\usepackage{float}
\usepackage{bm}
\usepackage{makecell}
\usepackage{empheq}
\usepackage{hyperref}
\hypersetup{
	colorlinks=true,
	linkcolor=blue,
	urlcolor=blue,
	citecolor=red
}
\usepackage{epsfig}%
\usepackage{pdfrender,xcolor}
\usepackage{cancel}

\usepackage{authblk}

\newcommand{\p}{\ensuremath{\partial{}}}

	\title{\boldmath \bf Dynamical systems analysis of tachyon dark energy model from a new perspective}
\bigskip

\author[1]{Saddam Hussain
	\thanks{\href{mailto:msaddam@iitk.ac.in}{msaddam@iitk.ac.in}}}
\author[2]{Saikat Chakraborty
	\thanks{\href{mailto:saikatnilch@gmail.com}{saikatnilch@gmail.com}\ (Corresponding Author)}}
\author[3]{Nandan Roy
	\thanks{\href{mailto:nandan.roy@mahidol.ac.th}{nandan.roy@mahidol.ac.th}\ (Corresponding Author)}}
\author[4]{Kaushik Bhattacharya
	\thanks{\href{mailto:kaushikb@iitk.ac.in}{kaushikb@iitk.ac.in}}}
\affil[1,4]{Department of Physics, Indian Institute of Technology, Kanpur\\ Uttar Pradesh 208016, India.}
\affil[2]{The Institute of Fundamental Study ``The Tah Poe Academia Institute'',\\ Naresuan University, Phitsanulok 65000, Thailand}
\affil[2]{Center for Space Research, North-West University,\\ Mahikeng 2745, South Africa}
\affil[3]{Centre for Theoretical Physics and Natural Philosophy, Mahidol University,\\ Nakhonsawan Campus, Phayuha Khiri, Nakhonsawan 60130, Thailand}

\begin{document}

\maketitle

\begin{abstract}
In this work, we present a new scheme to study the tachyon dark energy model using dynamical systems analysis by considering parametrization of the equation of state(EoS) of the dark energy. Both the canonical and phantom field dynamics are investigated. In our method, we do not require any explicit form of the tachyon potential. Instead of the potential, we start with an approximate form of the EoS of the tachyon field. This EoS is phenomenologically motivated and contains some dimensionless parameters. Using our method we can construct the dynamical system which gives rise to the time evolution of the universe. We have considered two different parametrizations of the EoS and studied the phase space dynamics in detail. Our analysis shows Taylor series parametrization of the EoS has serious cosmological limitations. We have also provided an example of how this method can be applied to coupled Tachyon models with a specific form of interaction. Our proposal is generic in nature and can be applied to other scalar field dark energy models.  
\end{abstract}

\section{Introduction}

The phenomenon of accelerated expansion of the universe ~\cite{SupernovaSearchTeam:1998fmf, SupernovaCosmologyProject:1998vns,  Meszaros:2002np, Planck:2014loa, ahn2012ninth} is still an unsolved mystery, even after two decades of its discovery. It is generally believed that there exists some mysterious component with negative pressure that is behind this accelerated expansion and it happens to be the dominant($70\%$) component of the universe. The cosmological constant ($\Lambda$) ~\cite{padmanabhan2006dark} with the constant equation of state (EoS) $(w_\lambda = -1)$ is the simplest and most successful model of the accelerated expansion of the universe. Because of it's success the \textit{$\Lambda$CDM} model in which the cosmological constant $\Lambda$ is considered as the candidate of the dark energy is given the status of the standard model of dark cosmology,  \textit{CDM} stands for cold dark matter. Although the \textit{$\Lambda$CDM} model is very successful in fitting the observed data, it seems that our universe is very highly fine-tuned to cause the cosmological constant to start dominating at a very specific time so that the universe can evolve into the present universe. This is one of the conceptual problems that the cosmological constant faces and is named as the cosmic coincidence problem. To explain the origin of the cosmological constant, another discrepancy arises between the theoretically predicted value of it and the observed value, which is of the order of hundreds of magnitude. Recently with the increment of our ability to constrain, the cosmological parameters with higher precision the \textit{$\Lambda$CDM} model also face challenges coming from the cosmological observations. The most important challenge at this moment is the Hubble($H_0$) tension. The early universe observations like  the CMB Planck collaboration \cite{Planck2020}, BAO \cite{BAO2017, BAO2011}, BBN \cite{BBN2021} and DES \cite{1DES2018, 2DES2018, krause2017dark} which consider the cosmological constant as the component of the dark energy estimated a value of the Hubble parameter to be $H_0 \sim (67.0 - 68.5)$ km/s/Mpc, which is lower than the observed value $H_0 = (74.03 \pm 1.42)$km/s/Mpc, obtained by observing the local universe using the distance ladder method from SH0ES \cite{Sh0ES2019} and H0LiCOW \cite{H0LiCOW2019} collaborations. The current discrepancy between these two types of data sets is of the order of $6 \sigma$. These difficulties with cosmological constants indicate clearly that there may be new physics involved in the late-time evolution of the universe and consequently an alternative to the cosmological constant should be investigated. In fact, there are recent claims that solving the $H_0$ tension issue does indeed require new late time physics \cite{Vagnozzi:2019ezj,Vagnozzi:2021tjv,Colgain:2022nlb,Aluri:2022hzs}.

Dynamical dark energy models in which the EoS of it varies with time have been proposed as alternatives to the cosmological constant. A wide variety of such dynamical dark energy models are abundant in the literature, a few of them are quintessence, k-essence, phantom, chaplygin gas, tachyon models, holographic DE models and so on \cite{sahni2006reconstructing, bamba2012dark,armendariz2001essentials, caldwell2002phantom, carroll2003can, kamenshchik2001alternative, sen2002tachyon, padmanabhan2002accelerated, copeland2006dynamics, amendola2010dark}. But there is no consensus on a particular model as each model has its benefits and drawbacks. 

The Quintessence scalar field model is the most popular and
well-studied dark energy model after the \textit{$\Lambda$CDM}
model. Recent observations have constrained the EoS of the dark energy
to be less than minus one. This fact has challenged the quintessence
model as with a canonical scalar field an EoS lower than minus one
cannot be achieved. There has been increasing interest in scalar field
dark energy with non-canonical kinetic term as for these models the
EoS can have values below minus one. The simplest one is the phantom
scalar field which has negative kinetic energy. In most general form
these models are known as k-essence models
\cite{armendariz2000dynamical} and
tachyon\cite{gibbons2002cosmological} is a special case of it with
Dirac–Born–Infeld (DBI) type of
action~\cite{chimento2004extended}. The concept of the tachyon field
is inspired by string theory where it arises naturally as a decay mode
of the D-branes~\cite{mazumdar2001assisted,sen2002tachyon,
  sen2002field}. Later it has been applied to cosmology and has been
extensively investigated as a candidate for dark
energy~\cite{cardenas2006tachyonic,choudhury2002cosmological,mughal2021multi,bagla2003cosmology,keresztes2009tachyon}. Recently
it has been also reported that distinguishing the tachyon dynamics
from quintessence scalar field is
difficult~\cite{ali2009transient,rajvanshi2021tachyonic}.

In this work we have considered the dark sector of the universe consists of the tachyon scalar field and a perfect barotropic fluid where the tachyonic sector is solely responsible for producing late time acceleration of the universe. Mathematically the dark energy sector is described by the Dirac-Born-Infeld (DBI) action. To make our analysis more general we have considered both the canonical tachyon field and the phantom tachyon scalar field by introducing a switch parameter $\epsilon$ in the action. According to  common practice some particular form of the tachyon potential is considered to study the dynamics of the tachyon field. One of the difficulties with the conventional approach is related to the arbitrariness of the choice of the potential. A wide variety of potentials can give rise to the same dynamics\cite{Roy:2018eug,Roy:2018nce} and as a result it is hard to distinguish the from the potential from observational results. In this work we have followed a different approach. Instead of choosing any specific tachyon potential we have chosen some approximate forms of the EoS of the tachyon field. These equations are time dependent and parametrized phenomenologically. As we know the primary dark energy component must have an equation of state with a value near $-1$ from observational inputs, we have assumed some workable forms of the EoS which in one case is a Taylor series like expansion of the EoS around the present time. In another case we have assumed an EoS of the tachyon field which closely resembles the EoS of dark energy when the Hubble parameter becomes approximately constant. In this paper we have worked with these two kinds of EoS for the tachyon field. Using these approximate forms of the EoS we have successfully generated the dynamics of the late time universe. Although the exact form of the potential of the scalar field is not necessary in our case, using our method one can approximately predict the form of the scalar field potential. The method presented here can be applied to models in which the tachyon field is coupled to matter. However, it cannot be readily applied to the general form of coupling considered in \cite{Gumjudpai:2005ry}. Nonetheless, a very similar interaction form, $\mathbb{Q} = Q \rho_{b} \dot{\phi} H$, can be analyzed using this method. The method which we have used is a very general one and can be used in other models of dynamical dark energy.   

The paper is presented as follows:  in Section 2, we discuss about the mathematical setup of the tachyon model and the construction of the autonomous system. Section 3 deals with dynamical systems analysis of the tachyon model with two different parametrizations of the EoS. This study includes both the quintessence and phantom field. Here we also discuss about some subtleties with our analysis. We extend our approach to a coupled tachyon model and studied the dynamics in Section 4. In section 5 we summarized and conclude our method and findings.

\section{Dynamics of the Tachyon Field}

Let us consider a homogeneous and isotropic universe described by the Friedmann-Lemaitre-Robertson-Walker(FLRW) metric
$$ds^2=-dt^2 + a^2(t)\,d{\bf x}^2\,,$$
where $a(t)$ is the scale-factor specifying the expansion of the universe. The dark energy sector is constituted by the tachyon field described by the Dirac-Born-Infeld (DBI) type of action which is given by:
\begin{equation}
\mathcal{S} = -\int V(\phi) \sqrt[]{ 1 + \epsilon \partial^\mu \phi \partial_\mu \phi} \ \sqrt[]{-g} d^4 x,
\end{equation}
where the parameter $\epsilon = \pm 1$. The plus sign is for the canonical tachyon field with positive kinetic energy and the minus sign is for the phantom type tachyon field which has negative kinetic energy. The total action of the system consists of the Einstein-Hilbert action for gravity and the action describing the behavior of a relativistic perfect fluid. In a homogeneous, isotropic and spatially flat universe,  evolution equations for the various dynamical variables can be written as:
\begin{subequations} \label{eqn_sys}
\begin{eqnarray} 
H^2 = \frac{\kappa^2}{3} \left(  \rho_b  + \frac{V}{\sqrt[]{1 - \epsilon \dot{\phi}^2}}\right), \label{H} \\
\dot{H} = - \frac{\kappa^2}{2} \left( ( \gamma_b -1)\rho_b +  \frac{\epsilon \dot{\phi}^2 V}{\sqrt[]{1 - \epsilon \dot{\phi}^2}}\right) , \label{Hdot}\\
\ddot{\phi} + 3 H \dot{\phi} (1 - \epsilon \dot{\phi}^2) + \epsilon \frac{V^\prime}{V} (1 - \epsilon \dot{\phi}^2) = 0, \label{phi}
\end{eqnarray} 
\end{subequations}
where $H=\frac{\dot{a}}{a}$ is the Hubble parameter and $\gamma$ is the equation of state parameter of the background field and $P_b = (\gamma_b -1) \rho_b$. In the above equations $\kappa^2 \equiv 8\pi G$, where $G$ is the universal gravitational constant. The dot represents a derivative with respect to cosmological time $t$. For  pressure less dust $\gamma_b = 1$ and for radiation $\gamma_b = 4/3$. Since we are interested in the late time dynamics of the universe from now onward we will consider the matter and tachyon field as the major components of the universe and neglect the contribution from radiation. Henceforth in this paper, we will always assume $\gamma_b = 1$. The pressure $(P_\phi)$ and density $(\rho_\phi)$ of the scalar field are given by
\begin{eqnarray}
P_\phi = - V(\phi) \ \sqrt[]{1 - \epsilon \dot{\phi}^2} ,\,\,\,\,\,\,\,\,\,\,
\rho_\phi = \frac{V(\phi)}{\sqrt[]{1 - \epsilon \dot{\phi}^2}}.
\end{eqnarray}
We will now cast the above equations as a set of constrained autonomous equations and then find out the nature of the critical points using dynamical systems analysis. 

In order to do so, we consider the following set of transformations:
\begin{equation} 
x = \dot{\phi}, \ y = \frac{\kappa\sqrt{V(\phi)}}{\sqrt{3} H}, \ \lambda = - \frac{V_{,\phi}}{\kappa V^{\frac{3}{2}}}, \ \Gamma = V \frac{V_{,\phi \phi}}{V_{,\phi} ^2},\; \sigma^2 = \dfrac{\kappa^2 \rho_b}{3 H^2}.
\label{dyn_var}
\end{equation}
One must note that in the present case the scalar field has inverse mass dimension and the potential has dimension of mass raised to the fourth power.
The subscript on a variable following a comma, as $V_{,\phi}$ represent derivative of $V$ with respect to $\phi$. Two subscripts after the comma imply two derivatives. These variables were first introduced in \cite{copeland2005needed} and later they were  used in \cite{ali2009transient,landim2015coupled}. The constrained equation from the Friedman equation, Eq.\eqref{H}, can be expressed in terms of dynamical variables as 
\begin{equation}\label{constraint_1}
	1 = \sigma^2 + \dfrac{y^2}{\sqrt{1 - \epsilon x^2}}.
\end{equation}
Using Eqs.\eqref{Hdot} and \eqref{phi}, the dynamical equations can be written as follows
\begin{subequations}\label{autonomous equation}
\begin{eqnarray} 
x^\prime &=& - (1 - \epsilon x^2) (3 x - \sqrt[]{3} \epsilon \lambda y), \label{x_prime} \\ 
y^\prime &=& \frac{y}{2} \left[ - \ \sqrt[]{3} \lambda x y - 3 \ {\sqrt[]{1 - \epsilon x^2}}\ y^2 + 3  \right], \label{y_prime} \\
\lambda^\prime &=& - \ \sqrt[]{3} \ \lambda^2 x y (\Gamma - \frac{3}{2}). \label{lam_prime}
\end{eqnarray}
\end{subequations}
In the above equations the prime represents derivatives with respect to $N\equiv\ln(a)$. The cosmological parameters of the scalar field like density parameter of the scalar field $\Omega_\phi$, equation of state parameter $ w_\phi$ and deceleration parameter $q$, the effective sound speed $ c_{s}^2 = P_{\phi, X}/\rho_{\phi, X}  $ \cite{Fei:2017fub}, where $X = -(1/2) g^{\mu \nu } \p_{\mu} \phi \p_{\nu} \phi $, can be written in terms of these new variables as:
\begin{eqnarray}
\Omega_\phi &=& \frac{y^2}{ \sqrt[]{1 - \epsilon x^2}}, \label{rho} \\
\gamma_\phi &\equiv& 1+ \frac{P_\phi}{\rho_\phi} = \epsilon \dot{\phi}^2 = \epsilon x^2 , \label{equation of state}\\
q &\equiv& -1-\frac{\dot{H}}{H^2} = - \left(1-\frac{3}{2}\gamma_b\right) - \frac{3}{2}y^2\frac{\gamma_b -\epsilon x^2}{\sqrt{1-\epsilon x^2}}, \label{decel_param}\\
\omega_{\rm tot} &\equiv& - \frac{2\dot{H}+3H^2}{3H^2} = \gamma_b-1 - y^{2}\frac{\gamma_b-\epsilon x^2}{\sqrt{1-\epsilon x^2}}, \label{omega_tot} \\
c_{s}^2 &= &  1 - \epsilon x^2\,. \label{sound_speed}
\end{eqnarray}
The EoS of the scalar field is; $\gamma_\phi=1+\omega_\phi$, where $\omega_\phi=P_\phi/\rho_\phi$. One can see from the set of autonomous equations in  Eq.\eqref{autonomous equation} that it is not closed unless one gives the information about the function $\Gamma$ or the form of the potential $V(\phi)$. Generally these equations can be closed in two different ways. One can consider a particular form of the potential and find out the corresponding $\Gamma$ from its definition or one can consider a particular form of $\Gamma$ and integrate back to get the corresponding class of potentials. This is the conventional approach used to solve such a set of autonomous equations. The difficulty arising from such a scheme is related to the fact that most of the time it becomes very difficult to guess the form of the potential or $\Gamma$ which will give rise to physically relevant behavior of the universe. One can use various forms of potential and predict the evolution of the late universe in various cases.

In the present paper we will not proceed in the conventional line. We will use a more phenomenological method to solve the autonomous system.  In our present work we start with some particular parametrization for the scalar field EoS ($\omega_\phi$). Using our prior knowledge from the recent cosmological observations\cite{Planck2020}, we expect the equation of state (EoS) for the dark sector will be around $-1$ or some preassigned value $\omega_0$ (close to $-1$). By considering a parametrization of the $\omega_\phi$ one will be able to find an expression of $\lambda$ in terms of $x$ and $y$, using the first two equations in the set of autonomous equations given in Eq.\eqref{autonomous equation}. For internal consistency we will use the form of $\lambda$ as obtained in the third autonomous equation and solve it. This solution will produce a form of the unspecified $\Gamma$ as a function of $x$ and $y$. In our approach all our ignorance about the functional form of the potential will be reflected in the functional form of $\Gamma$. As a result of this procedure the dimension of the phase space of our system will reduce from 3D to 2D as the variable $\lambda$ becomes redundant.

\section{Parametrization of equation of state of Dark energy } 

In the following section we discuss about the phase space dynamics of the tachyon scalar field model. We will consider two kinds of parametrization for $\omega_\phi$, the choice of these two parametrizations have different phenomenological motivations and can give rise to interesting cosmological dynamics. 
\subsection{First Parametrization}

In the first parametrization, the EoS of the scalar field is considered as: 
\begin{equation}\label{para01}
	\omega_{\phi}(N) = -1 + \omega_{1} N,
\end{equation}
where $\omega_{1}$ is a constant and can be constraint from  observations. As mentioned earlier, here $N=\log(a)$. The above mentioned parametrization has been inspired from Ref.\cite{Sangwan:2017kxi} where a more general form of it was considered: $ \omega_{\phi}(a) = \omega_{0}  - \omega^\prime \log (a)$. There $\omega^\prime$ is the derivative\footnote{Here the prime does not specify a derivative with respect to $N$.} of $\omega_\phi$ with respect to the scale-factor $a$, i.e., $ \omega^\prime = \frac{d \omega}{d a} \vert_{a=1} $.  The constraints on $\omega_{0}$ and $\omega^\prime$ came from $SNIa+BAO+ H(z) $ data and the constrained values were given as $  -1.09\le\omega_{0} \le -0.66 $ and $ -1.21 \le \omega^\prime \le 0.25 $. The constraint on  matter density was reported as $ 0.26 \le \Omega_{M} \le 0.32$. In our case we have taken $\omega_{0} =-1$ and $\omega_{1} = -\omega^\prime$. Using Eq.\eqref{equation of state} we can now write  $\gamma_\phi$ as:
\begin{subequations}\label{parametrized equation}
	\begin{eqnarray}
		\gamma_\phi  &=& 1 + \omega_{\phi} =  \omega_{1} N = \epsilon x^2\,. 
	\end{eqnarray}
\end{subequations}
Taking a derivative of Eq.\eqref{parametrized equation}, with respect to $N$, and equating the value of $x^\prime$ to the $x^\prime$ expression in Eq.\eqref{x_prime} we can derive another constraint equation as follows:
\begin{equation}\label{constraint_para01}
	-2 \epsilon x  (1 - \epsilon x^2) (3 x - \sqrt{3} \epsilon \lambda y)  = \omega_{1}\,.
\end{equation}
This constraint equation can be solved for $\lambda$   
\begin{equation}\label{new lam}
	\lambda = \frac{\frac{\omega_1}{1-x^2 \epsilon }+6 x^2 \epsilon }{2 \sqrt{3} x y }\,.
\end{equation}
To make the system of autonomous equations consistent the $\lambda^{\prime}$ equation in Eq.\eqref{lam_prime} should also be satisfied by the derivative of $\lambda$ given in Eq.\eqref{new lam} with respect to $N$. From the above mentioned consistency condition it is possible to express $\Gamma = f(x,y)$. To find the expression of $\Gamma$ one can see  appendix \ref{appen.1}. As result of this procedure the dynamical system now has essentially two independent variables, $x$ and $y$. The phase space has reduced from 3D to 2D.

One can now write down the effective autonomous equations in 2D. The new 2D autonomous system is given by: 
\begin{subequations}\label{ds_param1}
	\begin{eqnarray}
		x^\prime & =& \dfrac{\omega_{1}}{2 \epsilon x}\,, \\
		y^\prime & =& \frac{1}{2}y\left[- 3y^{2}\sqrt{1-\epsilon x^2} + 3(1-\epsilon x^2) - \frac{\omega_1}{2\left(1-\epsilon x^2\right)}\right]\,.
	\end{eqnarray}
\end{subequations}
One can notice that the system is symmetric under $x\mapsto-x$ and $y\mapsto-y$. One must note that by definition $y \ge 0$ and consequently the symmetry $y\mapsto-y$ is practically superficial although that transformation is a symmetry here. Next we study the phase space behavior of the canonical and phantom tachyon field for the current parametrization.

\subsubsection{Canonical tachyon (\texorpdfstring{$\epsilon= +1$}{Lg})}\label{subsec:canonical tachyon}

We first present the  dynamical system analysis for the case of a canonical tachyon field, where $\epsilon=1$. For this case, the phase space is compact because of the constraint in Eq.\eqref{constraint_1}. As a result we have:
\begin{equation}
  0\leq x\leq1, \qquad 0\leq y\leq(1-x^2)^{1/4}\,.
  \label{xyrangect}
\end{equation}
We have confined our attention to the first quadrant in the $x-y$ plane due to the symmetry $x\mapsto-x,\,y\mapsto-y$.  Here $y$ is always a positive semidefinite variable. It is interesting to note that the phase space is bounded by the curve
\begin{equation}\label{curve_1}
    y = (1-x^2)^{1/4},
\end{equation}
which is the locus of the points in the phase space that corresponds to an entirely scalar field dominated cosmology, except for the point $(x,y)=(1,0)$. This is because $\sigma=0$ at all the points on this curve except for $(x,y)=(1,0)$, as can be checked from the Eq.\eqref{constraint_1}. On the other hand, the line $y=0$ is the locus of the points in the phase space that corresponds to an entirely matter dominated cosmology, except for the point $(x,y)=(1,0)$. This is because $\sigma=1$ at all the points on this curve except for $(x,y)=(1,0)$.

We face a problem when we want to find the fixed points of the system. The autonomous system given in Eq.\eqref{ds_param1} diverges at $x=0,1$ and it seems like the only possible fixed point is at $(x \rightarrow \infty, y=0)$, which is outside the range specified in Eq.\eqref{xyrangect}. But this conclusion would be wrong because for an autonomous system of equations to qualify as a dynamical system, it needs to be at least first order differentiable \cite{Ellis,Coley:2003mj}, which is clearly violated by the system \eqref{ds_param1} at $x=0,1$. Therefore one should \textit{not} naively use the system in Eq.\eqref{ds_param1} to find the fixed points. This makes the analysis of the phase space behavior of the above mentioned system difficult. 

In reference \cite{Bouhmadi-Lopez:2016dzw} it has been suggested that a redefinition of the time coordinate can alter the form of the autonomous system of equations to make it differentiable, while keeping the phase space behavior intact. In the altered form the critical points of the system may appear inside the constrained phase space. Any redefinition of time alters the way the scale-factor evolves. In our analysis we are not directly working with cosmic time, we are working with $N=\log(a)$. A redefinition of time will produce a new $\bar{N}$. We define $\bar{N}$ as:  
\begin{equation}\label{time_redef_1}
    dN \rightarrow d\bar{N}=\frac{dN}{x(1-x^2)}\,.
\end{equation}
Sticking to the first quadrant ($x\ge 0,\,y\ge 0$) we see that $(d\bar{N}/dN)>0$, implying that $\bar{N}$ is always an increasing function of $N$ and thus our time redefinition is workable. Henceforth we will work with $\bar{N}$ and all the primes over $x$ and $y$ must be understood as derivatives of the respective variables with respect to $\bar{N}$. 

After the redefinition of the time variable, the dynamical system reduces to the following form
\begin{subequations}\label{ds_param1_eps+1}
	\begin{eqnarray}
		x' & =& \frac{\omega_{1}}{2}(1-x^2)\,, \\
		y' & =& \frac{1}{2}xy\left[-3y^{2}(1-x^2)^{3/2} + 3(1-x^2)^2 - \frac{\omega_1}{2}\right]\,.
	\end{eqnarray}
\end{subequations}
We emphasize that, the time redefinition in Eq.\eqref{time_redef_1}, and consequently the regularized dynamical system in Eq.\eqref{ds_param1_eps+1}, is valid only in the first quadrant. In a forthcoming subsection we will explicitly show the method using which one may construct the dynamics in the second quadrant if one knows the dynamics in the first quadrant. Here we analyze the system in the first quadrant. There is only one fixed point $(x,y)=(1,0)$. Since this fixed point lies at the intersection of the curve specified in Eq.\eqref{curve_1} and the line $y=0$, we cannot definitively determine the cosmology corresponding to this fixed point. 
\begin{table}[H]
	\centering
	\begin{tabular}{|c|c|c|c|c|}
		\hline
		Points & $ (x,y) $ & Eigenvalues & Stability & Cosmology \\
		\hline
		$ P_{1} $ & $ (1,0) $ & $ \left\lbrace -\omega_{1},-\frac{\omega_1}{4} \right\rbrace $ & \multicolumn{1}{|p{4cm}|}{\centering Stable for $\omega_1>0$ \\ Unstable for $\omega_1<0$} & $a(t)\sim t^{2/3}$ \\
		\hline
		\end{tabular}
	\caption{The nature of one of the critical points for the first parametrization corresponding to $\epsilon = +1$. The cosmological evolution corresponding to the fixed point can be derived from the expression of the deceleration parameter in Eq.\eqref{decel_param}.}
	\label{tab:para_01_eps_pl_nature}
\end{table}
The nature of the critical point is tabulated in Tab.\ref{tab:para_01_eps_pl_nature}. This critical point is stable (unstable) for $\omega_{1}>0$ ($\omega_{1}<0$) and non-hyperbolic for the special case of $\omega_{1}=0$, i.e. cosmological constant. The cosmology corresponding to this critical point is dominated by the kinetic part of the scalar field and near it the scale-factor is evolving as $a(t)\sim t^{2/3}$.
\begin{figure}[H]
	\begin{minipage}[b]{0.5\linewidth}
		\centering
		\includegraphics[scale=.4]{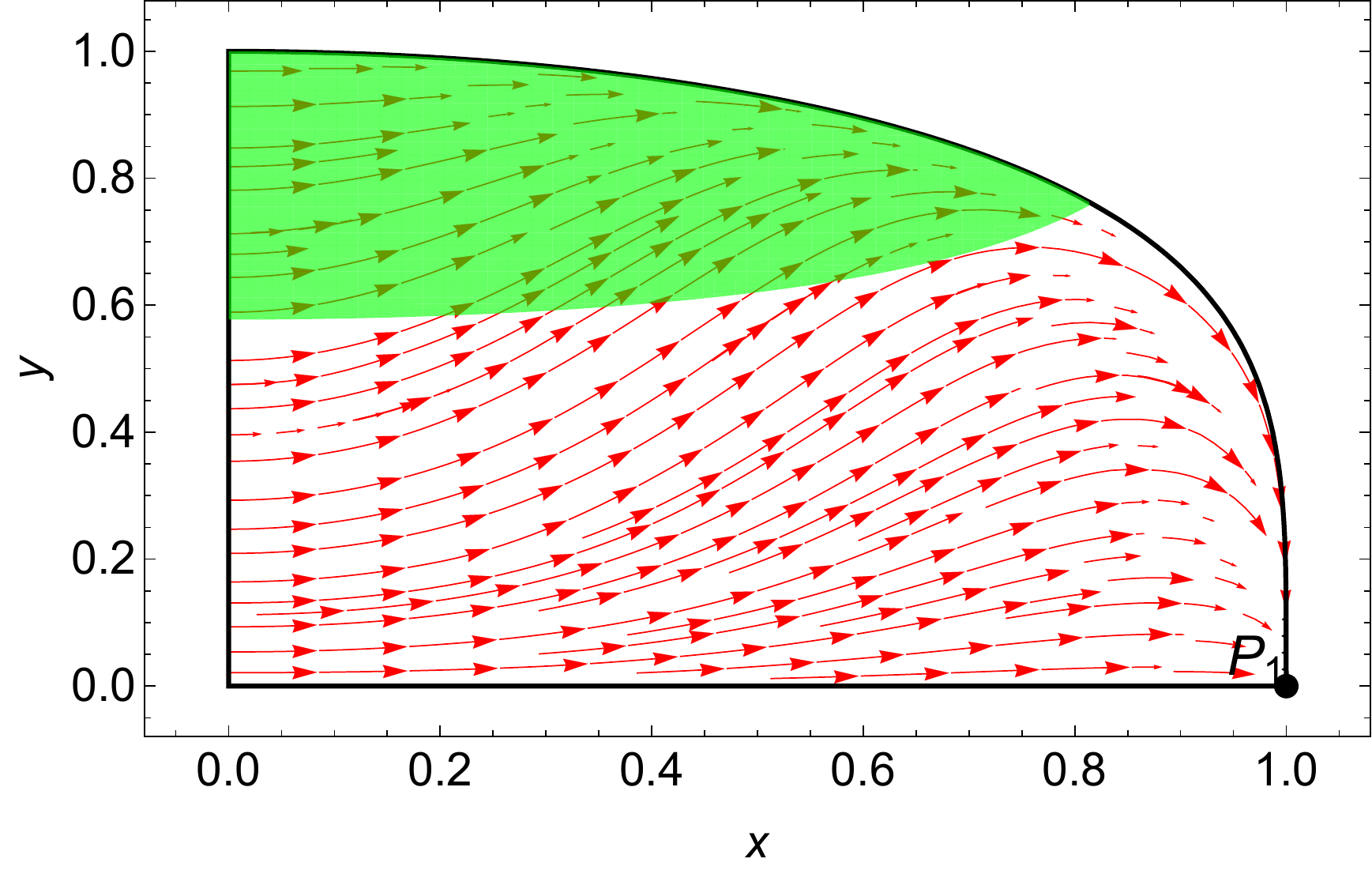}
		\subcaption{The phase space plot for $\epsilon= +1, \omega_{1} =0.3	$.}
		\label{fig:ist_para_plus_eps}
	\end{minipage}
	\hspace{0.2cm}
	\begin{minipage}[b]{0.5\linewidth}
		\centering
		\includegraphics[scale=.4]{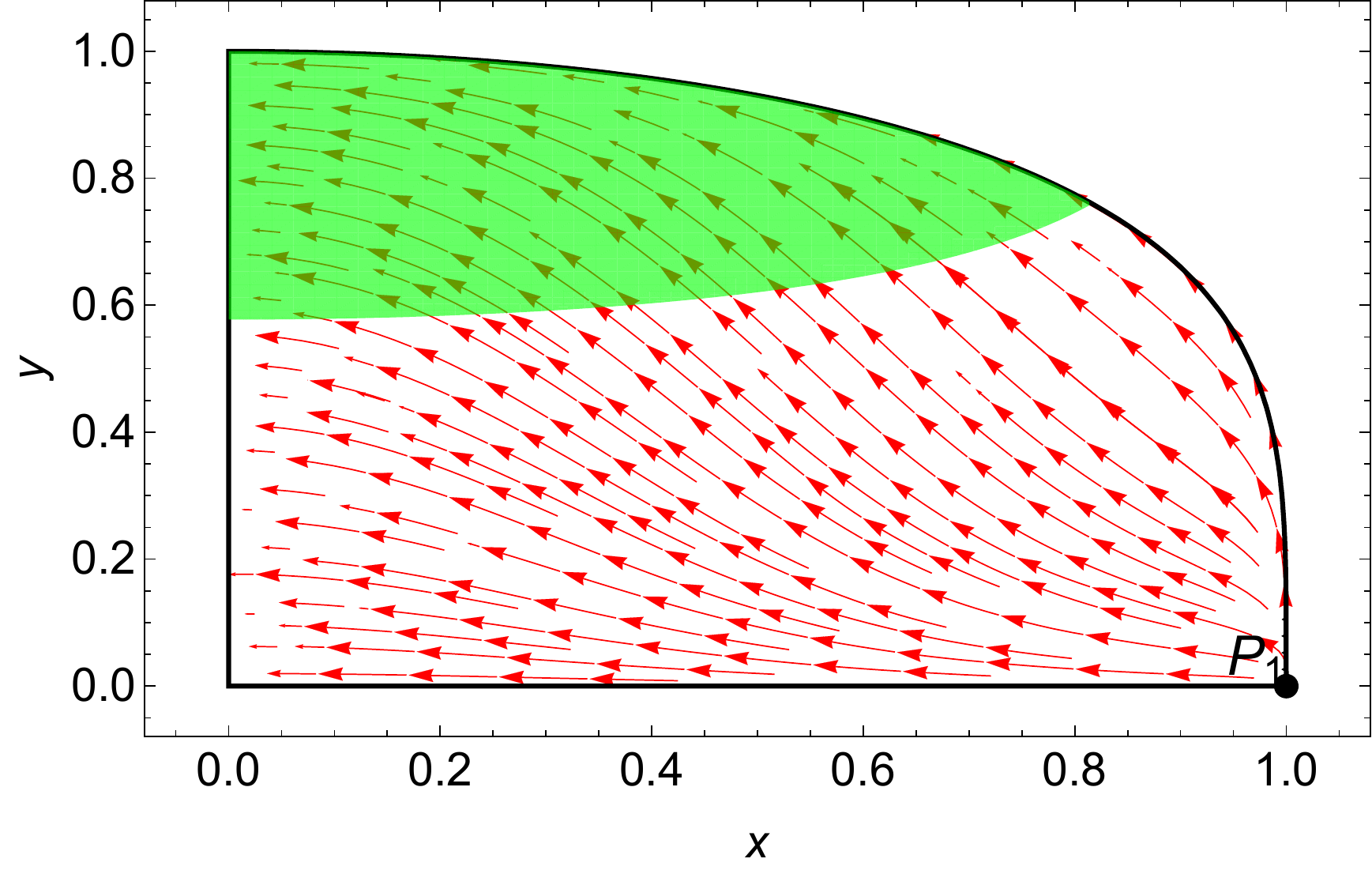}
		\subcaption{The phase space plot for $\epsilon= +1, \omega_{1} =-0.3	$ . }
		\label{fig:ist_para_plus_eps_momega}
	\end{minipage}
	\caption{The compact phase space for canonical tachyon with the parametrization given in Eq.\eqref{para01}. The phase space is compactified by the constraint Eq.\eqref{constraint_1}. At the present epoch $N=0$, $x=0$ (see Eq.\eqref{parametrized equation}). The green region represents an accelerated phase with $ -1 \le \omega_{\rm tot} < -1/3 $ (obtained using Eq.\eqref{omega_tot}). The condition for the absence of gradient instability, namely $ 0\le c_{s}^2 \le 1 $ is satisfied on the entire phase space.}
\end{figure}


The nature of the phase space plots around the point $x=1$, $y=0$ is shown in Fig.\ref{fig:ist_para_plus_eps} and Fig.\ref{fig:ist_para_plus_eps_momega}. The plots clearly show the stability issue related to this fixed point. Corresponding to this redefined autonomous system we see that $\Gamma$ becomes $ -1 $ for this critical point which shows that at the late time phase $\Gamma$ becomes constant. 


\subsubsection{Phantom tachyon (\texorpdfstring{$\epsilon=-1$}{Lg})}\label{subsec:phantom tachyon}

Here we present the results related to the phantom tachyon. For this case, $\epsilon=-1$ and as a consequence one can see from Eq.\eqref{constraint_1} that the phase space is not compact. We first try to compactify the phase space as in that case only we can exhaustively show phase space dynamics. In the following we compactify the phase space using the following prescription:
\begin{equation}\label{dyn_var_comp}
    X=\frac{x}{\sqrt{1+x^2}}, \qquad Y=\frac{y}{\sqrt{1+y^2}}.
\end{equation}
In terms of $X,Y$ the Friedmann constraint in Eq.\eqref{constraint_1} becomes
\begin{equation}\label{constraint_2}
1 = \sigma^2 + \sqrt{1-X^2}\,\frac{Y^2}{1-Y^2}\,.    
\end{equation}
The phase space in the $(X,Y)$ coordinates is compact, because, from the constraint equation Eq.\eqref{constraint_2},
\begin{equation}
    0\leq X\leq1, \qquad 0\leq Y\leq \frac{1}{(1+\sqrt{1-X^2})^{1/2}},
\end{equation}
where, again, we have confined our attention to the first quadrant because of the symmetry $x\mapsto-x,\,y\mapsto-y$. Similar to the previous case the phase space is bounded by the curve
\begin{equation}\label{curve_2}
    Y = \frac{1}{(1+\sqrt{1-X^2})^{1/2}},
\end{equation}
which is the locus of the points in the phase space that corresponds to an entirely scalar field dominated cosmology, except for the point $(X,Y)=(1,1)$. This is because $\sigma=0$ at all the points on this curve except for $(X,Y)=(1,1)$, as can be checked from the Eq.\eqref{constraint_2}. On the other hand, the lines $X=1$ and $Y=0$ are the locus of the points in the phase space that corresponds to an entirely matter dominated cosmology, except for the point $(X,Y)=(1,1)$. This is because $\sigma=1$ at all the points on this line except for $(X,Y)=(1,1)$.

\begin{figure}[H]
	\begin{minipage}[t]{0.5\linewidth}
		\centering
		\includegraphics[scale=.4]{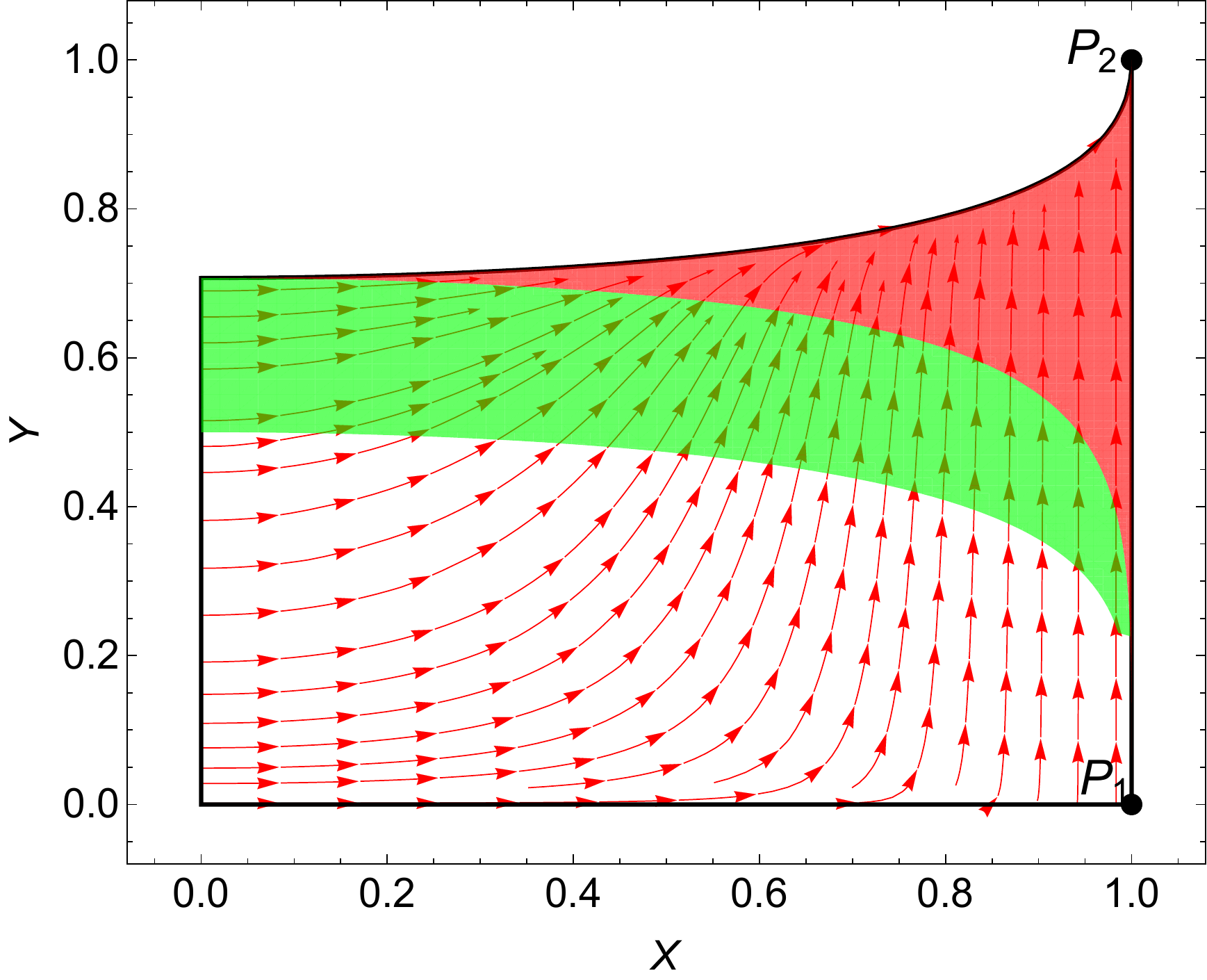}
		\subcaption{The phase space plot for $\epsilon= -1, \omega_{1} =-0.5	$.}
		\label{fig:01 para eps m1}
	\end{minipage}
	\hspace{0.2cm}
	\begin{minipage}[t]{0.5\linewidth}
		\centering
		\includegraphics[scale=.4]{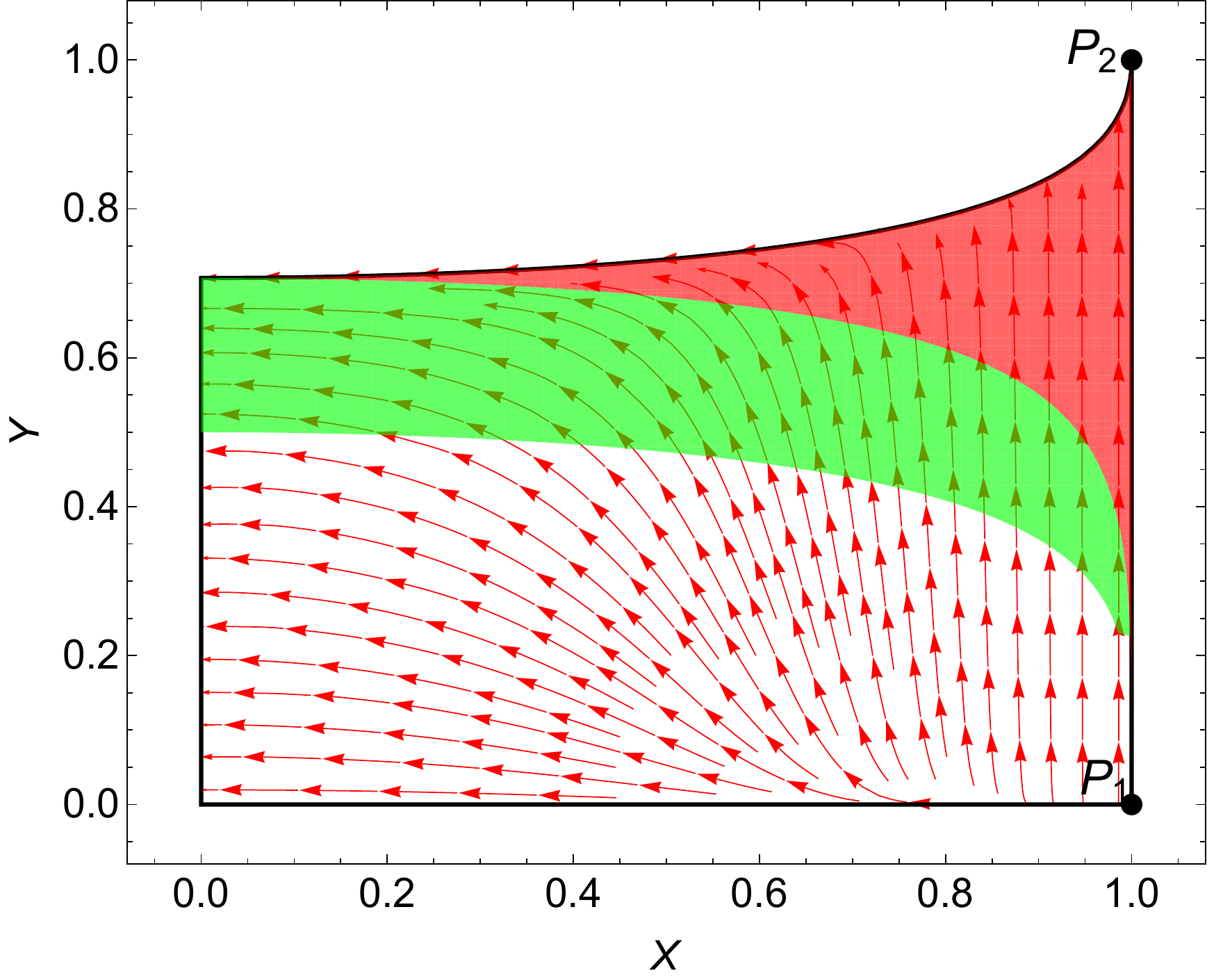}
		\subcaption{The phase space plot for $\epsilon= -1, \omega_{1} =0.5	$ . }
		\label{fig:ist para_0.5}
	\end{minipage}
\caption{The compact phase space for phantom tachyon with the parametrization given in  Eq.\eqref{para01}. The phase space is compactified by the constraint Eq.\eqref{constraint_2}. At the present epoch $N=0$, $x=0$. The green region represents an accelerated phase with $ -1 \le \omega_{\rm tot} < -1/3 $. The red region represents the phantom phase with $\omega_{\rm tol} < -1$.}
\end{figure}

The effective 2D autonomous dynamical system is described by the following equations:
	\begin{subequations}
	    \begin{eqnarray}
		X' & =& -\frac{\omega_{1}}{2X}(1-X^2)^2 \,,\\
		Y' & =& \frac{1}{2}Y\left[- \frac{3Y^2}{\sqrt{1-X^2}} + \frac{3(1-Y^2)}{1-X^2} - \frac{\omega_1}{2}(1-X^2)(1-Y^2)\right]\,.
	\end{eqnarray}
	\end{subequations}
Here the primes designate differentiation with respect to $N=\log(a)$. From the above equations, it is seen that even after the compactification of the phase space the system diverges at $X=0,1$. To tackle this problem we have adopted a similar strategy as adopted in the previous  case. We redefine the time coordinate such that
\begin{equation}\label{time_redef_2}
    dN \rightarrow d\bar{N}=\frac{dN}{X(1-X^2)}.
\end{equation}
Using $\bar{N}$ we can express the dynamical system in a new form as:
\begin{subequations}\label{ds_param1_eps-1}
	\begin{eqnarray}
	    X' & =& -\frac{\omega_{1}}{2}(1-X^2)^{3}\,, \\
		Y' & =& \frac{1}{2}X Y\left[- 3Y^{2}\sqrt{1-X^2} + 3(1-Y^2) - \frac{\omega_1}{2}(1-X^2)^{2}(1-Y^2)\right]\,,
	\end{eqnarray}
\end{subequations}
where the derivatives (primes) are now with respect to $\bar{N}$. We emphasize that the time redefinition in Eq.\eqref{time_redef_2}, and consequently the regularized dynamical system in Eq.\eqref{ds_param1_eps-1}, is valid only in the first quadrant. In the next subsection we will explicitly show the method using which one may construct the dynamics in the second quadrant if one knows the dynamics in the first quadrant. Here we present the dynamics of the system only in the first quadrant. There are two fixed points $(X,Y)=(1,0),(1,1)$ in the first quadrant. From the constraint in Eq.\eqref{constraint_2} it is clear that at the fixed point $(1,0)$, $\sigma=1$, i.e. the universe near this fixed point is in a matter dominated phase. The fixed point $(1,1)$ lies at the intersection of the curve specified in Eq.\eqref{curve_2} and the line $X=1$. Consequently the cosmology corresponding to this fixed point cannot be determined definitively.
\begin{table}[H]
	\centering
	\begin{tabular}{|c|c|c|c|c|}
		\hline
		Points & $ (X,Y) $ & Eigenvalues & Stability & Cosmology \\
		\hline
		$ P_{1} $ & $ (1,0) $ & $ (0,3/2) $ & \multicolumn{1}{|p{4cm}|}{\centering Unstable for $\omega_1>0$ \\ Saddle for $\omega_1<0$} & Matter dominated \\
		\hline
		$ P_{2} $ & $ (1,1) $ & $ (0,-3) $ & \multicolumn{1}{|p{4cm}|}{\centering Saddle for $\omega_1>0$ \\ Stable for $\omega_1<0$} & Indeterminate \\
		\hline
\end{tabular}
\caption{The nature of critical point for first parametrization corresponds to $\epsilon = -1$. }
\label{tab:epslm1}
\end{table}
The nature of the fixed points are tabulated in Tab.\ref{tab:epslm1}. Note that the fixed points in this case are non-hyperbolic so that their stability cannot be inferred from a Jacobian analysis. Instead, their stability can be inferred investigating the nature of the various invariant submanifolds of the system. The mathematical analysis is presented in appendix \ref{app:stability}. The phase space plots showing the nature of dynamical evolution around the fixed points are shown in Fig.\ref{fig:01 para eps m1} and Fig.\ref{fig:ist para_0.5}. The two plots correspond to two values of $\omega_1$, in one case $\omega_1>0$ and in the other case $\omega_1<0$.


\subsubsection{Phase dynamics in the second quadrant}

In the above discussion we have focused our attention only to the first quadrant $x>0,\,y>0$, because the system in Eq.\eqref{ds_param1} possesses the symmetry $x\mapsto-x$ and $y\mapsto-y$. The time redefinitions in Eqs.\eqref{time_redef_1},\eqref{time_redef_2} and the regularized dynamical systems in Eqs.\eqref{ds_param1_eps+1},\eqref{ds_param1_eps-1} are valid in the first quadrant only. Consequently the Figs.\ref{fig:ist_para_plus_eps},\ref{fig:ist_para_plus_eps_momega},\ref{fig:01 para eps m1},\ref{fig:ist para_0.5} all shows the phase space dynamics in the first quadrant. As we argued before, the symmetry $y\mapsto-y$ is practically superficial because by definition $y\ge 0$. One may, however, ask what would be the phase space dynamics in the second quadrant $x<0,\,y>0$. All possible cosmological scenarios with $\dot{\phi}<0$ constitutes the second quadrant. If we focus our attention to the second quadrant, the time redefinitions in Eqs.\eqref{time_redef_1},\eqref{time_redef_2} should be modified as:
\begin{equation}
    dN \rightarrow d\bar{N}=\frac{dN}{-x(1-x^2)}
\end{equation}
and
\begin{equation}
    dN \rightarrow d\bar{N}=\frac{dN}{-X(1-X^2)}.
\end{equation}
respectively. This is because one needs to respect the condition $(d\bar{N}/dN)>0$ to preserve the arrow of time. The regularized dynamical systems in Eqs.\eqref{ds_param1_eps+1},\eqref{ds_param1_eps-1} would then be modified only by an overall minus sign in front. Consequently the phase space dynamics in the second quadrant will just be a reflection of that in the first quadrant against the line $x=0$ or $X=0$.

In fact, this can be also argued as follows. The dynamical system in Eq.\eqref{ds_param1} is of the form
\begin{equation}
    u_x = f_x(x,y)\,, \qquad u_y = f_y(x,y)\,.
\end{equation}
where $u_{x,y}$ are the Cartesian components of the ``phase flow'' at each point $(x,y)$. The functions $f_{x,y}(x,y)$ are such that $f_x(-x,y)=-f_x(x,y)$ and $f_y(-x,y)=f_y(x,y)$. As one goes from first to second coordinate, the $x$-component of the flow velocity is inverted whereas the $y$-component remains intact. As a result of these observations we see the phase space dynamics in the first quadrant is given by Figs.\ref{fig:ist_para_plus_eps} or \ref{fig:ist_para_plus_eps_momega} for the canonical tachyon and Figs.\ref{fig:01 para eps m1} or \ref{fig:ist para_0.5} for the phantom tachyon. The phase space dynamics in the second quadrant will be a reflection of these plots against $x=0$.

We also notice that the first and second quadrant is completely disjoint. There is no direct way to ``glue'' the phase portraits of the first quadrant with corresponding portraits in the second quadrant. This is related to the time redefinitions we have used and consequently the regularized dynamical systems are valid in only either the first quadrant or the second quadrant. The physical interpretation of this fact is that the time derivative of the field, $\dot{\phi}$, cannot change sign at any point during the  cosmic evolution. One can also interpret this in another way. Note that the  nature of the fixed points remains the same irrespective of whether one considers the first quadrant or the second quadrant. As a result one may conclude that the actual sign of $\dot{\phi}$ does not really matter as the cosmological dynamics is concerned. Whether it is positive or negative, the cosmological dynamics remains the same.


\subsubsection{Analytical solution for the asymptotic behavior}\label{subsec:analytic}

For the particular parametrization that we have considered here it is possible to analytically confirm the results that we have found from the phase space analysis. The parametrization essentially fixes the solution for $x(N)$ from Eq.\eqref{parametrized equation}
\begin{equation}
    \epsilon x^{2}(N) = \omega_1 N.
\end{equation}
The equation for $y(N)$ is then
\begin{equation}
    \frac{dy(N)}{dN} = \frac{1}{2}y \left[-3y^{2}\sqrt{1-\omega_{1}N} + 3(1-\omega_{1}N) - \frac{\omega_{1}}{2(1-\omega_{1}N)}\right].
\end{equation}
Let us now find out the asymptotic behavior for canonical and phantom tachyon.

\begin{itemize}
\item {\bf For canonical tachyon} ($\epsilon=1$), and $x^{2}(N) = \omega_1 N$: For $\omega_{1}>0$, the solution is well-defined in the future up to a finite e-folding $N=N_{\rm max}=1/\omega_1$ but undefined in the past. For $\omega_{1}<0$, the solution is well-defined in the past up to a finite e-folding $N=N_{\rm min}=1/\omega_1$ but undefined in the future. The finite value $N_{\rm max,min}$ comes from demanding $1-\omega_{1}N>0$, which is required for $\frac{dy(N)}{dN}$ to be real. We emphasize that, a finite value for the e-folding $N$ corresponds to a finite value of the scale factor $a(t)$, but does \textit{not} necessarily correspond to a finite value of the cosmological time $t$. As $N\rightarrow N_{\rm max,min}$, $x\rightarrow1$ and
    \begin{equation}
        \frac{dy(N)}{dN} \sim \frac{y}{4\left(N-\frac{1}{\omega_1}\right)}.
    \end{equation}
    Integrating the above we get, 
    \begin{equation}
        y(N) \sim \bigg\vert N-\frac{1}{\omega_1} \bigg\vert.
    \end{equation}
    Therefore $y(N)\rightarrow0$ as $N\rightarrow N_{\rm max,min}$. This is consistent with the results obtained in the subsection \ref{subsec:canonical tachyon} as the fixed point $P_1$, which is a future attractor for $\omega_{1}>0$ and a past attractor for $\omega_1<0$, has the coordinates $(x,y)=(1,0)$.

\item {\bf For a phantom tachyon} ($\epsilon=-1$), and $x^{2}(N) = -\omega_1 N$: For $\omega_{1}>0$, the solution is well-defined in the past but undefined in the future. For $\omega_{1}<0$, the solution is well-defined in the future but undefined in the past. As $N\rightarrow\pm\infty$, $x\rightarrow\infty$ ($X\rightarrow1$) and
    \begin{equation}
        \frac{dy(N)}{dN} \sim -\frac{3}{2}y\omega_{1}N.
    \end{equation}
    Integrating the above we get, 
    \begin{equation}
        y(N) \sim e^{-\frac{3}{4}\omega_{1}N^2}.
    \end{equation}
    Therefore $y(N)\rightarrow0$ ($Y\rightarrow0$) as $N\rightarrow-\infty$ for $\omega_1>0$ and $y(N)\rightarrow\infty$ ($Y\rightarrow1$) as $N\rightarrow\infty$ for $\omega_1<0$. This is consistent with the results obtained in the subsection \ref{subsec:phantom tachyon} as the fixed point $P_1\equiv(X,Y)=(1,0)$ is a past attractor for $\omega_1>0$ and $P_2\equiv(X,Y)=(1,1)$ is a future attractor for $\omega_1<0$.
\end{itemize}


\subsubsection{Some comments about Taylor series approximation}
\label{subsec:taylor_param}

The parametrization that we have worked with in this section, namely the one given by Eq.\eqref{para01}, is the first order Taylor series approximation of $\omega_{\phi}(N)$ around the present epoch $N=0$. The coefficients of the Taylor expansion can be constrained via observations. As a first approximation we have assumed that the first order Taylor series approximation of $\omega_{\phi}(N)$ remains valid for the entire domain of consideration. One can take into account higher order terms in the Taylor series to get a better approximation for $\omega_{\phi}(N)$. Nonetheless, our analysis with the first order Taylor series approximation suffices to establish the generic methodology to work with Taylor series parametrizations. For example, consider the parametrization containing the second order Taylor series approximation
\begin{equation}\label{para02}
    \omega_{\phi} = -1 + \omega_{1}N + \omega_{2}N^2 \,,
\end{equation}
which gives 
\begin{equation}\label{parametrized equation 2}
    \epsilon x^2 = \omega_{1}N + \omega_{2}N^2 \,.
\end{equation}
For the first order Taylor series parametrization in Eq.\eqref{para01} we have taken the derivative of Eq.\eqref{parametrized equation} once with respect to $N$ and then used the dynamical equation of $x$ from Eq.\eqref{autonomous equation} to arrive at the constraint in Eq.\eqref{constraint_para01}. For the second order Taylor series parametrization given in Eq.\eqref{para02} we need to take the derivative of Eq.\eqref{parametrized equation} twice with respect to $N$ and use the dynamical equations in Eq.\eqref{autonomous equation} in place of $x',\,y',\,\lambda'$ to arrive at a similar constraint equation. In general, if one takes the $n$-th order Taylor series approximation as the parametrization for $\omega_{\phi}(N)$, one needs to take derivatives with respect to $N$, $n$ times, and use the dynamical equations in Eq.\eqref{autonomous equation} at each step to replace of $x',\,y',\,\lambda'$.

Although the Taylor series parametrization has a straightforward motivation, there is a serious drawback with this parametrization as long as tachyonic dark energy is concerned. As we have seen with the first order Taylor series parametrization in subsection \ref{subsec:analytic}, the solution $x(N)$ is undefined either in past ($N<0$) or future ($N>0$). This is not an unique result valid only for the first order parametrization. Even for higher order Taylor series parametrizations, the solution will always be undefined either in past or future. As a consequence of this the Taylor series parametrizations of any order can be used to describe the cosmological dynamics either from the matter dominated epoch in the past up to the present epoch or from the present epoch up to some future asymptotic time, but \textit{not} an entire dynamics starting from the matter dominated epoch in the past through the present day to a future asymptotic. One can infer that Taylor series parametrizations are not really compatible with tachyonic dark energy models.


\subsection{Second Parametrization}\label{sec:2nd_param}

The issues with the Taylor series parametrizations that we have pointed out in subsection \ref{subsec:taylor_param} motivates us to try other parametrizations for the equations of state of tachyonic dark energy. Here we consider the parametrization studied in \cite{Usmani:2008ce, Chevallier:2000qy}. This is a completely different way of parametrizing a time dependent EoS of the scalar field. The parametrization is given as:
\begin{equation}\label{2nd_para}
	\omega_{\phi} = \omega_{0} + \omega_{1} (t \dot{H} / H)\,.
\end{equation}
In this parametrization $\omega_{0}$ and $\omega_{1}$ are dimensionless constant parameters. In this parametrization $\omega_{\phi} \to \omega_0$ in a pure dark energy dominated phase, where $\dot{H}\sim 0$. The factor $t/H$ is included before $\dot{H}$ to make $\omega_1$ dimensionless.  
This parametrization can be associated with the field equations via the equation of state, $\omega_{\phi}=-1+\epsilon x^2 $. The last equation can also be written as $ 1 + \omega_{0} + \omega_{1} (t \dot{H} / H) = \epsilon x^2 $. Taking the derivative of this equation with respect to $ t $ and then converting the derivatives in terms of $N$, for the case where $\omega=0$, one gets a relation as:
\begin{equation}\label{general 3rd auto_x}
x' =\dfrac{1}{2 \epsilon - (\epsilon x^2 -1 - \omega_{0}) (2/x^2 + \epsilon/(1-\epsilon x^2)) } \bigg[ \dfrac{3}{2} \dfrac{\epsilon x y^2}{\sqrt{1- \epsilon x^2}} (-\omega_1 + \epsilon x^2 -1 - \omega_{0} ) - \sqrt{3} \lambda y (\epsilon x^2 - 1 - \omega_{0})\bigg]\,.
\end{equation}
Equating this expression of $ x $ to the value of $x^{\prime}$ in Eq.\eqref{autonomous equation} one can find out a constraint for the the variable $\lambda$ as: 
\begin{equation}\label{new_lambda}
	\lambda = \frac{\frac{3 x y^2 \epsilon  \left(x^2 \epsilon -\omega_{0}-\omega_{1}-1\right)}{2 \sqrt{1-x^2 \epsilon } \left[2 \epsilon -\left(\frac{\epsilon }{1-x^2 \epsilon }+\frac{2}{x^2}\right) \left(x^2 \epsilon -\omega_{0}-1\right)\right]}-3 x \left(x^2 \epsilon -1\right)}{\frac{\sqrt{3} y \left(x^2 \epsilon -\omega_{0}-1\right)}{2 \epsilon -\left(\frac{\epsilon }{1-x^2 \epsilon }+\frac{2}{x^2}\right) \left(x^2 \epsilon -\omega_{0}-1\right)}-\sqrt{3} y \epsilon  \left(x^2 \epsilon -1\right)}\,.
\end{equation}
In this dynamical system $\lambda$ is related to the derivative of the potentials. In our approach we are not working with the exact form of the scalar field potential, we are using an extra condition consistently which reduces the phase space dimension by one.
If we take the derivative of this $\lambda$ in Eq.~\eqref{new_lambda} with respect to $N$  and equate the resultant expression with the one appearing in  Eq.~\eqref{lam_prime}, we will get the form of $ \Gamma $ in terms of $ x $ and $ y $. One can find the evolution of $ \Gamma $ in terms of $ x $ and $ y $. In this way, Eq.~\eqref{new_lambda} provides another constraint on the system which reduces the phase space dimension from 3D to 2D.

\subsubsection{Canonical tachyon (\texorpdfstring{$\epsilon =+ 1$}{Lg}) }

For the normal tachyon field, the expression of $\lambda$ from Eq.~\eqref{new_lambda} becomes: 
\begin{equation}\label{}
	\lambda = -\frac{\sqrt{3} \left(-4 \sqrt{1-x^2} x (\omega_{0}+1)+x^5 \left(2 \sqrt{1-x^2}-y^2\right)+x^3 \left(2 \sqrt{1-x^2} (\omega_{0}+1)+y^2 (\omega_{0}+\omega_{1}+1)\right)\right)}{4 \left(1-x^2\right)^{3/2} y (\omega_{0}+1)}\, . 
\end{equation}
Here we can see that $\lambda$ becomes singular at $ x= \pm 1, y =0 $ and $\omega_{0} = -1$.  Differentiating $ \lambda $ with respect to $ N $, gives: 
\begin{multline}\label{}
	\lambda' =  \frac{1}{4 \left(1-x^2\right)^{5/2} y^2 (\omega_{0}+1)}  \sqrt{3}  \bigg[2 x \left(1-x^2\right)^{3/2} \left(x^4+x^2 (\omega_{0}+1)-2 (\omega_{0}+1)\right) y' \\
	+x^3 \left(x^2-1\right) y^2 y' \left(-x^2+\omega_{0}+\omega_{1}+1\right)-x^2 y^3 x' \left(2 x^4-5 x^2+3 (\omega_{0}+\omega_{1}+1)\right)+ \\
	2 \sqrt{1-x^2} y \left(3 x^6+x^4 (\omega_{0}-4)-x^2 (\omega_{0}+1)+2 (\omega_{0}+1)\right) x' \bigg]\,.
\end{multline}
Equating the above equation with Eq.~\eqref{lam_prime} produces the expression of $ \Gamma $. The expression of $\Gamma$ is presented in appendix \ref{app:norm_tachyon_2nd_para}.

Using the above techniques the phase space dimension reduces by one and the autonomous equations of the effective system can be expressed as:
\begin{subequations}
	\begin{eqnarray}
		x' & = & -\frac{3 x^3 \left[x^2 \left(2 \sqrt{1-x^2}-y^2\right)-2 \sqrt{1-x^2} (\omega_{0}+1)+y^2 (\omega_{0}+\omega_{1}+1)\right]}{4 \sqrt{1-x^2} (\omega_{0}+1)} \, , \label{auto_x_3rd_para}\\
		y' & =&  \frac{1}{2} y \bigg[-3 \sqrt{1-x^2} y^2+ \dfrac{1}{4 \left(1-x^2\right)^{3/2} (\omega_{0}+1)} 3 x \bigg(-4 \sqrt{1-x^2} x (\omega_{0}+1)+ 	\label{auto_y_3rd_para} \nonumber\\
		& & x^5 \left(2 \sqrt{1-x^2}-y^2\right) +x^3 \left(2 \sqrt{1-x^2} (\omega_{0}+1)+y^2 (\omega_{0}+\omega_{1}+1)\right)\bigg) \bigg]\,.
\end{eqnarray}
\end{subequations}
Although $y$ is by definition only defined in the positive branch, i.e., it is always positive in our case (with zero minimum), we see that the above set of equations have the symmetry $x \mapsto -x$ and $y \mapsto -y$. This symmetry is reflected in the table of fixed points. We tabulate all the fixed points of the system although physically only the positive values of $y$ are significant. We have found the fixed points of the above set of autonomous equations and these critical points are tabulated in Tab.\ref{tab:critical_point_3rd_para}.
\begin{table}[t]
\centering
\begin{tabular}{cccc}
		\hline 
		\multicolumn{4}{c}{Critical points for  $ \epsilon = +1$. }\\
		\hline 
		Points & $ x $ & $ y $ & Eigenvalues ($ E_{1} $ \&  $ E_{2} $)\\
		\hline
		$ P_{0} $ & 0 & 0 & $ (0,3/2) $\\
		\hline
		$ P_{1,2} $ & $ \mp \sqrt{1 + \omega_{0}} $  & 0 & \makecell{$- 3(1 + \omega_{0}), -3 \omega_{0}/2 $}\\ 
		\hline 
		$ P_{3,4} $ & $  - \sqrt{1 + \omega_{0} - \omega_{1}} $ & $ \mp  (-\omega_{0} + \omega_{1}) ^{1/4}$ & \makecell{$ \dfrac{-3}{2(1+ \omega_{0})}(1+ \omega_{0} - \omega_{1})^2 $, \\
			 $  3 (\omega_{0} - \omega_{1}) $. }\\
		\hline

		$ P_{5,6} $ & $  \sqrt{1 + \omega_{0} - \omega_{1}} $ &  $ \mp  (-\omega_{0} + \omega_{1}) ^{1/4}$  & same as above \\
		\hline		
		$ P_{7,8} $	& 0 & $ \mp 1 $ & \makecell{$ (-3,0) $} \\
		\hline
	\end{tabular}
\caption{The critical points in the second parametrization. }
\label{tab:critical_point_3rd_para}
\end{table}
In the present case, there are eight critical points which are dependent on the values of $\omega_{0}$ and $\omega_{1}$. To find the stability of each point, we linearize the autonomous system around the fixed point and find the Jacobian matrix. The eigenvalues of the Jacobian matrix are $ E_{1} $ and $ E_{2} $. Based on the sign of the real coefficients of the eigenvalues the stability of the system is determined. The eigenvalues of the system have been tabulated in Tab.\ref{tab:critical_point_3rd_para}. A fixed point is stable (unstable) if all the eigenvalues have real negative(positive)  parts. For alternate signs of the real parts of eigenvalues, the point becomes a saddle point. If any eigenvalue becomes zero for a fixed point, then the system's stability will no longer be determined by linearization. In the last case, one can employ either the center manifold theorem or find it numerically by solving the differential equation. In order to understand the nature of the points the total EoS, $\omega_{\rm tot}$, and sound speed, $ c_{s}^2 $, of the system have been shown in Tab.\ref{tab: 3rd_para_eos_sound_lambda}. 
\begin{table}[h!]
	\centering
	\begin{tabular}{ccccccc}
		\hline
		Points & $ (x,y) $	&$\omega_{\rm tot}$ & $ c_{s}^2 $ & $ \Gamma $ & $\lambda$ & Stability\\
		\hline
		$ P_{0} $ & $ (0,0) $	&$ 0 $ & 1 & $ \infty $ & $ \infty $ & Unstable \\
		\hline
		$ P_{1,2} $& $ (\mp 0.32, 0) $ & 0 &  $ - \omega_{0}$  & $ \dfrac{1}{2 \omega_{0} +2}+1 $  & $ \infty $ & Saddle\\
		\hline
		$ P_{3,4} $& $ (-0.39, \mp 0.96) $	 & $\omega_{0}$  $-\omega_{1}$ & $ -\omega_{0} + \omega_{1} $  & $ 3/2 $&  $\pm  \frac{\sqrt{3} \sqrt{\omega_{0}-\omega_{1}+1}}{\sqrt[4]{\omega_{1}-\omega_{0}}} $ & Stable \\
		\hline
		$ P_{5,6} $& $ (0.39, \mp 0.96) $ & $\omega_{0}$  $-\omega_{1}$ & $ -\omega_{0} + \omega_{1} $ & $ 3/2 $ & $\mp \frac{\sqrt{3} \sqrt{\omega_{0}-\omega_{1}+1}}{\sqrt[4]{\omega_{1}-\omega_{0}}}$& Stable\\
		\hline
		$ P_{7,8} $& $ (0, \pm 1) $ & $ -1 $ & $ 1 $ & $ \infty $ & 0 & Stable\\
		\hline
	\end{tabular}
\caption{The critical points and its nature for $\epsilon = 1$, corresponding to $\omega_{0} = -0.9$ and $\omega_{1} = -0.05$.} 
\label{tab: 3rd_para_eos_sound_lambda}
\end{table}

Out of the eight critical points, two critical points are only $ \omega_{0} $ dependent and these are denoted by $ P_{1,2} $. These critical points exists for $ \omega_{0}\ge - 1$. The eigenvalues for these points show that they are saddle points for $-1< \omega_{0} <0  $ and stable points for $ \omega_{0}>0 $. These points share $\omega_{\rm tot} = 0$, signifies the matter phase of the universe. From the sound speed limit  $ 0\le c_{s}^2 \le 1 $, we infer that $\omega_{0}$ must take negative values. 

There is a set of critical points, consisting of points $ P_{7,8} $, where the points are constants. For this set, one of the eigenvalues becomes zero, for each element, and hence the stability of the points cannot be determined in a conventional way. To find the stability of these critical points, we used the numerical technique and found that the points are stable. The EoS and sound speed around these points are $ -1 $ and $ 1 $ respectively. 

The rest of the critical points from $ P_{3} $ to $ P_{6} $ are $\omega_{0}$ and $\omega_{1}$ dependent. In order to have real critical points $ (\omega_{0} - \omega_{1}) \ge -1 $. For stable fixed points one requires $ \omega_{0} < \omega_{1} $. In order to produce the accelerating solution we require $ \omega_{\rm tot} < -1/3 $ and from the sound speed condition we find that $ \omega_{\rm tot} = -c_{s}^2 $. From this we observe that as $\omega_{\rm tot}$ approaches $ -1 $,  sound speed becomes $ 1 $. Hence we may infer that by taking negative values of $\omega_{0}$ and $ \omega_{1} $ one may obtain the desired late time dynamics of the universe. Hence for further analysis we have chosen $\omega_{0} = -0.9$ and $\omega_{1} = -0.05 $. 
\begin{figure}[H]
	\centering
	\includegraphics[scale=0.6]{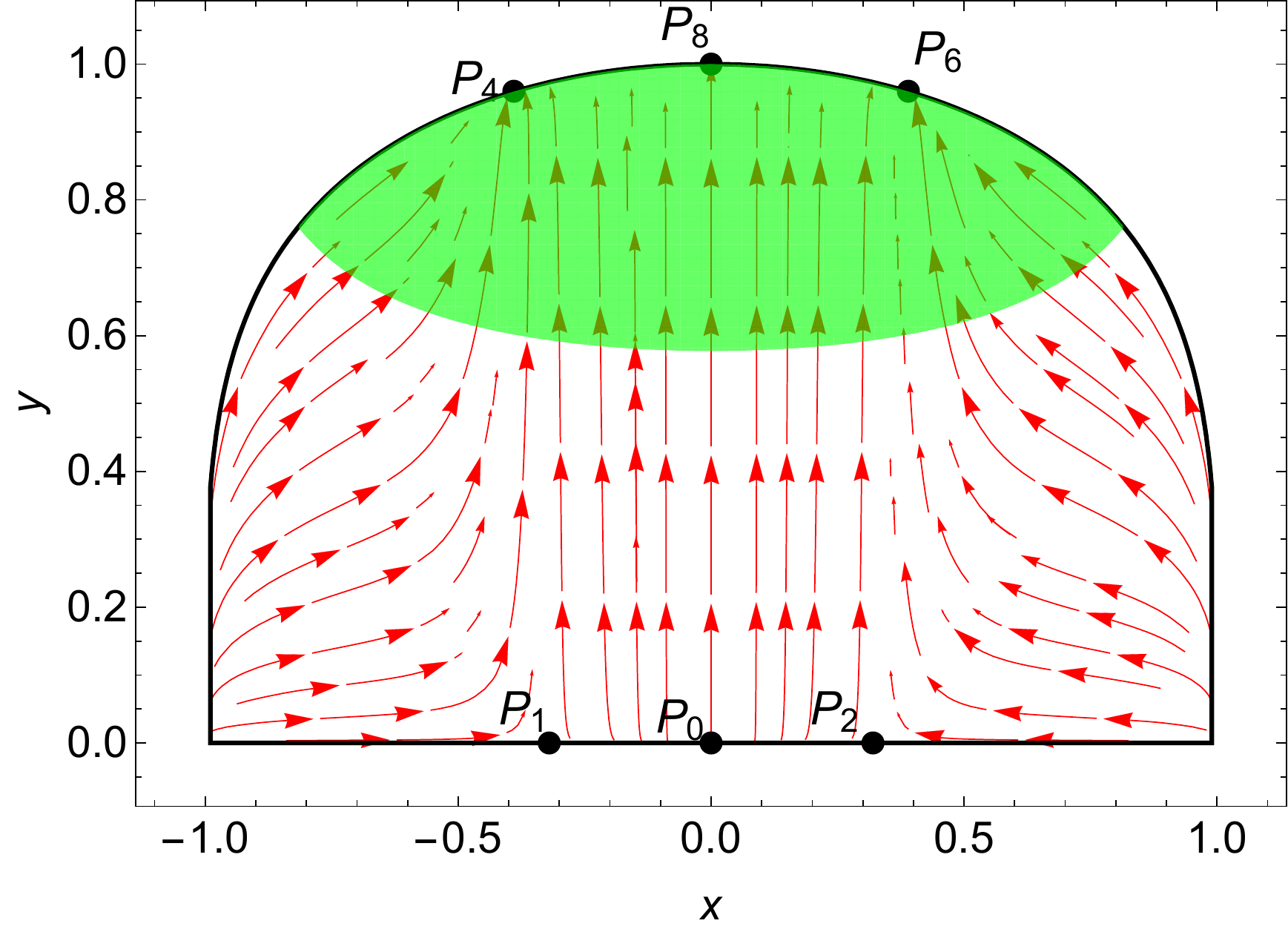}
	\caption{The phase space plot for $\epsilon= 1, \, \omega_{0} =-0.9	$ and $\omega_{1} = - 0.05$. The phase space is constrained from Eq.~\eqref{H}. In the green region $ -1 \le \omega_{\rm tot} < -1/3 $.}
	\label{fig: 3rd_para_phase_space_eps_1}
\end{figure}

The phase portrait  of the system has been shown in Fig.\ref{fig: 3rd_para_phase_space_eps_1}. There are two separate regions in the phase plot. The green regions represent the accelerated expansion phase of the universe whereas the white region represents decelerated expansion phase of the universe. One can clearly see all the solutions for the considered values of $\omega_0 =  -0.9$ and $\omega_1 = -0.05$ specifying decelerated expansion phase are on the $y=0$ line. This implies deep in the matter-dominated era the potential of the scalar field was close to zero and after that the potential becomes nontrivial, giving rise to the dynamics of the scalar field. The phase space is symmetric around both the $x$ and $y$ axes. We have only plotted the physically significant region. The sound speed limit remains between \(0\) to \(1\) in the entire region.   
The trajectories are attracted towards  $P_4, P_6, P_8$. All these points are scalar field dominated fixed points. From  Tab.\ref{tab: 3rd_para_eos_sound_lambda} these fixed points are stable in nature and act as late time attractors. 

\begin{figure}[H]
	\centering
	\includegraphics[scale=0.5]{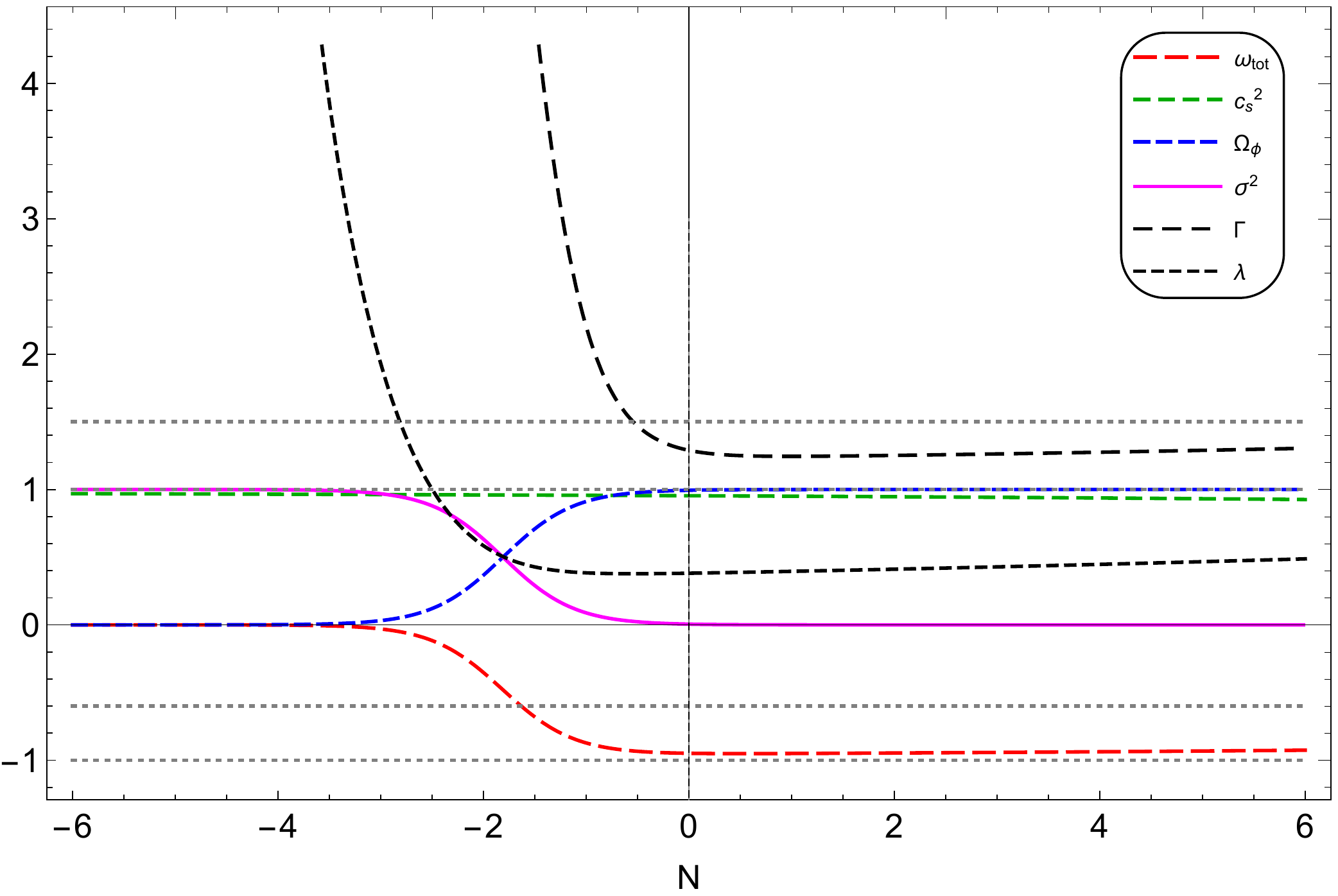}
	\caption{The evolution plot for $\epsilon= 1,\, \omega_{0} =-0.9	$ and $\omega_{1} = - 0.05$.  }
	\label{fig: 3rd_para_evo_eps_1}
\end{figure}

We have plotted the system variables such as $ \omega_{\rm tot}, c_{s}^2, \Omega_\phi, \sigma^2, \Gamma $ and $ \lambda $ in Fig.\ref{fig: 3rd_para_evo_eps_1} with respect to $N $ by solving the autonomous equations for $ x' $ and $ y'$. In the early epoch of the universe, the EoS of the system starts from $ 0 $ when the fluid energy density parameter $\sigma^2$ dominates over the scalar field energy density parameter. At this epoch, the sound speed is nearly $ 1 $, and both $ \lambda $ and $ \Gamma $ are significantly large. As the universe evolves, the EoS decreases towards the negative value, and at some point where the tachyon field energy density has increased significantly, the EoS becomes saturated to $ -0.85 $.  In the last phase the parameters $ \lambda $ saturates at $ < 1.3 $ whereas $ \Gamma $ saturates at \(0.4\). The system's late time EoS can go very close to $ -1 $ depending on the choice of $ \omega_{0} $ and $ \omega_{1} $. This signifies that the non-phantom tachyon field can describe the late time acceleration with significant sound speed.  

Although in our formalism one does not need the exact form of the potential, as the form of the EoS of the scalar field modifies the autonomous equations and produces the desired dynamics of the system, one may approximately find out the form of the potential using the phase space dynamics. Here we present an approximate scheme using which the form of $V(\phi)$ can be found out. The method discussed here is a general one and can be applied to most of the cases discussed in this paper. As because the functional form of the potential is not required in our case we have not calculated the potential in all other cases. If one is really interested to know the approximate form of the potential then the method discussed below can always be used.   
To find out the functional form of the potential we will use the definition $ \lambda = - \dfrac{V_{,\phi}}{\kappa V^{3/2}} $. In the present case $\lambda$ is a function of $ x $ and $ y $ as given in Eq.~\eqref{new_lambda}. From the autonomous equation, we have found the evolution of $ x $ and $ y $ in terms of  $ N=\log a $. We can express $ V(\phi ) $ in terms of dynamical variable as,
$$\kappa	\int \lambda  \mathrm{d} \phi   = - \int \dfrac{\rm{d} V}{V^{3/2}}\,,$$
which yields
\begin{eqnarray}
  \left[ \kappa /2 \int \lambda \, \frac{x}{H} \,  \rm{d}(\log a) \right] ^{-2} & = & V(\phi)\,.
  \label{potential form}
\end{eqnarray}
\begin{figure}[H]
	\centering
	\includegraphics[scale=0.8]{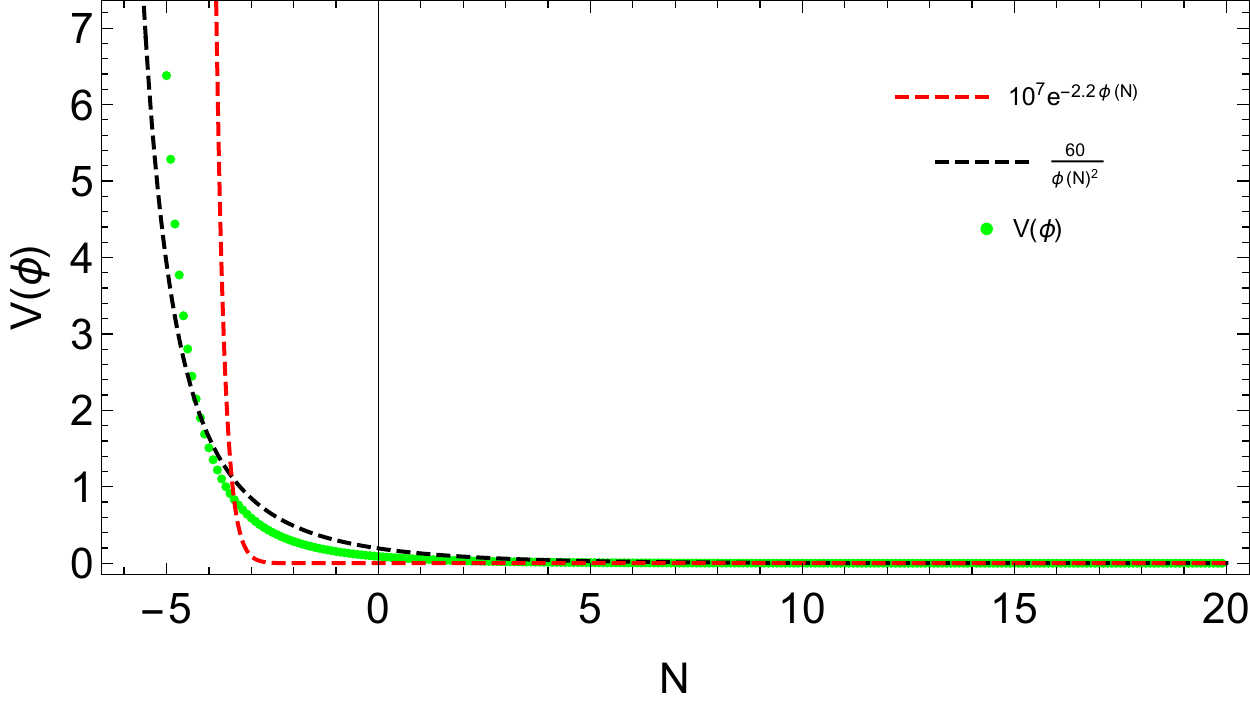}
	\caption{The exact potential and its approximations for $\epsilon= 1, \omega_{0} =-0.9	$ and $\omega_{1} = - 0.05$.}
	\label{fig:pot}
\end{figure}

\noindent In the above equation $H$ can be expressed in terms of the
phase space variables $x$ and $y$\footnote{In this regard one must note that one can also use the the definition of $\Gamma(N)$ to determine the functional form of the potential as we know the numerical values of $\Gamma$ for all relevant values of $N$. In this case one will have to solve a second order differential equation for $V(\phi)$ for each $N$ as the expression of $\Gamma$ contains $V_{,\phi\phi}$. This shows that one can in principle determine the approximate functional form of the potential in multiple ways.}.  We can then numerically integrate
and find the form of the potential of the system, expressed as a function of $N$. In
Fig.\ref{fig:pot} we have found the potential $ V(\phi) $, which is
evolving with $ \log a $, by assuming $ \kappa=1 $. The exact form of
the potential obtained from this process is shown in green color. We
see that at the early epoch the potential of the tachyon field is
relatively very high and as the system evolves the potential saturates
at a lower value. The potential becomes considerably low at the late
time phase of the universe. One can approximately find the functional form of the potential. Here we present two possible forms. It is seen that $V(\phi)=60/\phi^2$ nicely fits the actual green curve\footnote{One can easily find out how this potential evolves with $N$ as we know how $V(\phi)$ depends on $\phi$ and $\phi=\phi(N)$.}. This kind of a potential often arises in k-essence models. There is another possible form of the potential given by $V(\phi)=10^7\times e^{-2.2\phi}$. This form of the potential also closely matches with the exact value of $V(\phi)$ in the late phase of the universe whereas in the very early phase this approximation does not work. This form of potential is generally used in quitessence models. These two forms of the approximate potential show that our analysis can predict a functional form of $V(\phi)$ if required. Henceforth we will not calculate the functional form of $V(\phi)$ as this form is not required to predict the dynamics of our system.  

\subsubsection{Phantom tachyon (\texorpdfstring{$\epsilon=-1$}{Lg})}

After the normal tachyons we will now deal with the phantom tachyons. From the first Friedmann equation in Eq.~\eqref{constraint_1} it becomes apparently clear that phase space is not constrained as: 
\begin{equation}\label{}
	0 \le y^2 \le \sqrt{1 + x^2}\,.
\end{equation}
Since $ -\infty <x < \infty $ we have  $ 0 < y < \infty $. As a result of this we will first like to compactify the phase space.  To compactify the phase space we introduce two variables $ (X,Y) $ \cite{messias2011dynamics} defined as:
\begin{equation}\label{poincare_constrained}
	X= \dfrac{x}{\sqrt{1+ x^2 + y^2}}, \quad Y= \dfrac{y}{\sqrt{1+ x^2 + y^2}}\,.
\end{equation}
Using these variables one can observe that our phase space is constrained as we get 
\begin{equation}\label{}
	0 \le Y^{2} \le \dfrac{1-X^2}{2-X^2}\,, 
\end{equation}
where $-1\le X \le 1$. One can now express Eq.~\eqref{general 3rd auto_x}, Eq.~\eqref{y_prime} and Eq.~\eqref{new_lambda} in terms of the new variables $ X  $ and $ Y$. The expression of $\lambda$ and $\Lambda$ for the present case are given in appendix \ref{app:phan_tach_lamb_second}. In this case, we have nine critical points of the system which are tabulated in Tab.\ref{tab:poincare_critical_points_epsm1}. In order to have real critical points and positive sound speed $\omega_{0}$ must be negative and less than $ \omega_{1}$, hence we shall analyze the system for $\omega_{0} = -3, \, \omega_{1} = 1/2$. 
\begin{table}[t!]
	\centering
	\begin{tabular}{ccccccc}
		\hline
		\multicolumn{7}{c}{Critical points in terms of compact variables  }\\
				\hline
		Points & $ X $ & $ Y $& $\omega_{\rm tot}$ & $ c_{s}^2 $ & $ \Gamma $   & $\lambda$ \\
		\hline
		$ P_{0} $ & 0 & 0& 0 & $1 $ & $\infty$ & $\infty$\\
		\hline 
		$ P_{1,2} $ & $ \mp \frac{\sqrt{\omega_{0}+1}}{\sqrt{\omega_{0}}} $ & 0 & 0 & $ -\omega_{0} $ & $ \frac{3 + 2 \omega_{0}}{2(1+ \omega_{0})} $ & $\infty$\\
		\hline 
		$ P_{3,4} $ & $ -\frac{\sqrt{\sqrt{\omega_{1}-\omega_{0}}+\omega_{0}-\omega_{1}}}{\sqrt{\omega_{0}-\omega_{1}}} $ & $ \mp \frac{\sqrt{(\omega_{0}-1) \sqrt{\omega_{1}-\omega_{0}}-2 \omega_{0}+\omega_{1}}}{\sqrt{-(\omega_{0}-\omega_{1}+1) \left(\sqrt{\omega_{1}-\omega_{0}}+\omega_{0}\right)}} $ & $ \omega_{0} - \omega_{1} $ & $-\omega_{0} + \omega_{1}$ & $ 3/2  $ & see Fig.\ref{fig:lambda_poincare_fixed_point_contour}\\
		\hline 
		$ P_{5,6} $ & $ \frac{\sqrt{\sqrt{\omega_{1}-\omega_{0}}+\omega_{0}-\omega_{1}}}{\sqrt{\omega_{0}-\omega_{1}}} $ & $ \mp \frac{\sqrt{(\omega_{0}-1) \sqrt{\omega_{1}-\omega_{0}}-2 \omega_{0}+\omega_{1}}}{\sqrt{-(\omega_{0}-\omega_{1}+1) \left(\sqrt{\omega_{1}-\omega_{0}}+\omega_{0}\right)}} $ & $ \omega_{0} - \omega_{1} $ & $-\omega_{0} + \omega_{1}$ & $ 3/2  $ & see Fig.\ref{fig:lambda_poincare_fixed_point_contour}\\
		\hline 
		$ P_{7,8} $ & 0 & $ \pm 1/\sqrt{2} $ & $ -1 $ & $ 1 $ & $\infty$ & $ 0 $\\
		\hline
 	\end{tabular}
 \caption{The critical points and corresponding physical parameters for $\epsilon =-1$. }
 \label{tab:poincare_critical_points_epsm1}
\end{table}
\begin{figure}[H]
	\centering
	\includegraphics[scale=0.6]{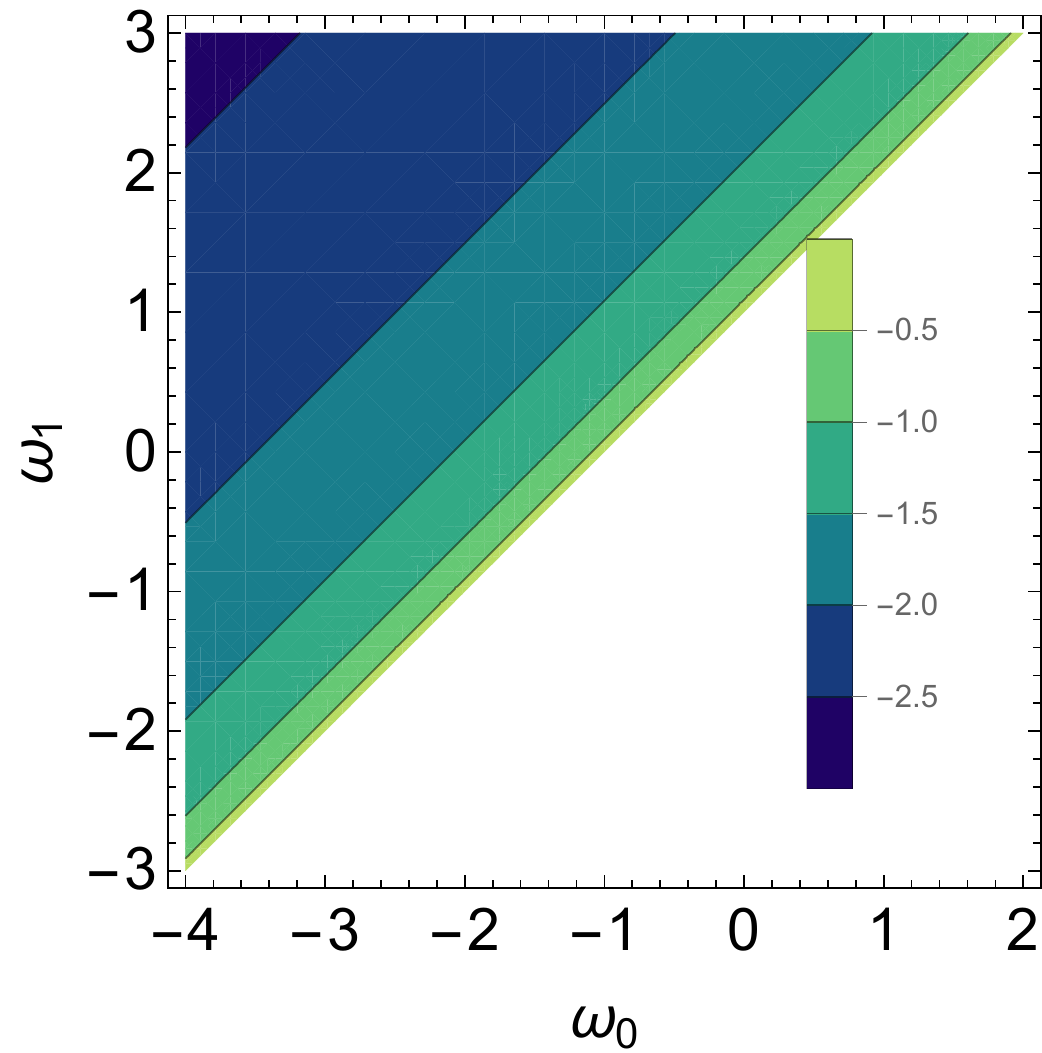}
	\caption{The values of $\lambda$ corresponds to the fixed points $ P_{3} $ to $ P_{6} $ is $\omega_{0}$ and $\omega_{1}$ dependent. Here $\lambda$ is negative if both $ (X, Y) $ have the same sign and positive for the alternate sign (keeping $|X|, |Y|$ the same) while the magnitude of $\lambda$ remains the same in both cases.}
	\label{fig:lambda_poincare_fixed_point_contour}
\end{figure}

We have shown all the critical points of the system in the present case, these critical points include both positive and negative values of $Y$. Although only the critical points with positive values of $Y$ matters. We have shown the full set of critical points to specify the symmetry of the problem. The value of $\lambda$ for the critical points $P_3$ to $P_6$ can be found from
the plot in Fig.\ref{fig:lambda_poincare_fixed_point_contour}. The plot shows a contour where in the white region there exists no real, finite value of $\lambda$. The colour codes give the range of $\lambda$ values which one may expect if one chooses appropriate $(\omega_0,\omega_1)$ pairs. 

In Tab.\ref{crit_points_phantom_tach_2} we have represented the critical points for $\omega_0=-3$
and $\omega_1=1/2$ for $\epsilon=-1$. The table shows the values of various parameters of the model for the specific values of $\omega_0,\omega_1,\epsilon$. One can see that in this present case one can have superluminal sound propagation. This effect is particular to this model and one cannot modify this result. In the present case as $\omega_{\rm tot}=-c_s^2$ we will always have $c_s^2>1$ for the phantom case $\omega_{\rm tot} < -1$. Lately, it is known that various kinds of k-essence theories can in principle have superluminal sound propagation. This propagation happens only in the medium with a specific configuration of scalar fields and not in vacuum. In a certain way, Lorentz invariance is not lost because of the particular medium produced by the scalar field. 
\begin{table}[H]
	\centering
	\begin{tabular}{ccccc}
		\hline
		Points & $ (X,Y) $ & $ \omega_{\rm tot} $ & $ c_{s}^2 $& Stability \\
		\hline 
		$ P_{0} $ & $ (0,0) $ & $ 0 $ & $ 1 $ & Unstable \\
		\hline
		$ P_{1,2} $ & $ (\mp \sqrt{2/3} , 0) $ & $ 0 $ &$ 3 $ & Unstable \\
		\hline 
		$ P_{3,4} $ & $ (-0.68, \mp 0.59) $ & $ -7/2 $ &$ 7/2  $ & Saddle\\
		\hline 
		$ P_{5,6} $ & $ (0.68, \mp 0.59) $ & $ -7/2 $ & $ 7/2 $ & Saddle\\
		\hline 
		$ P_{7,8} $ & $ (0, \mp 1/\sqrt{2}) $ & $ -1 $ &$  1$ & Stable\\
		\hline	
	\end{tabular}
	\caption{Critical points for $\omega_{0} = -3, \omega_{1}= 1/2$. }
        \label{crit_points_phantom_tach_2}
\end{table}
\begin{figure}[H]
	\centering
	\includegraphics[scale=0.6]{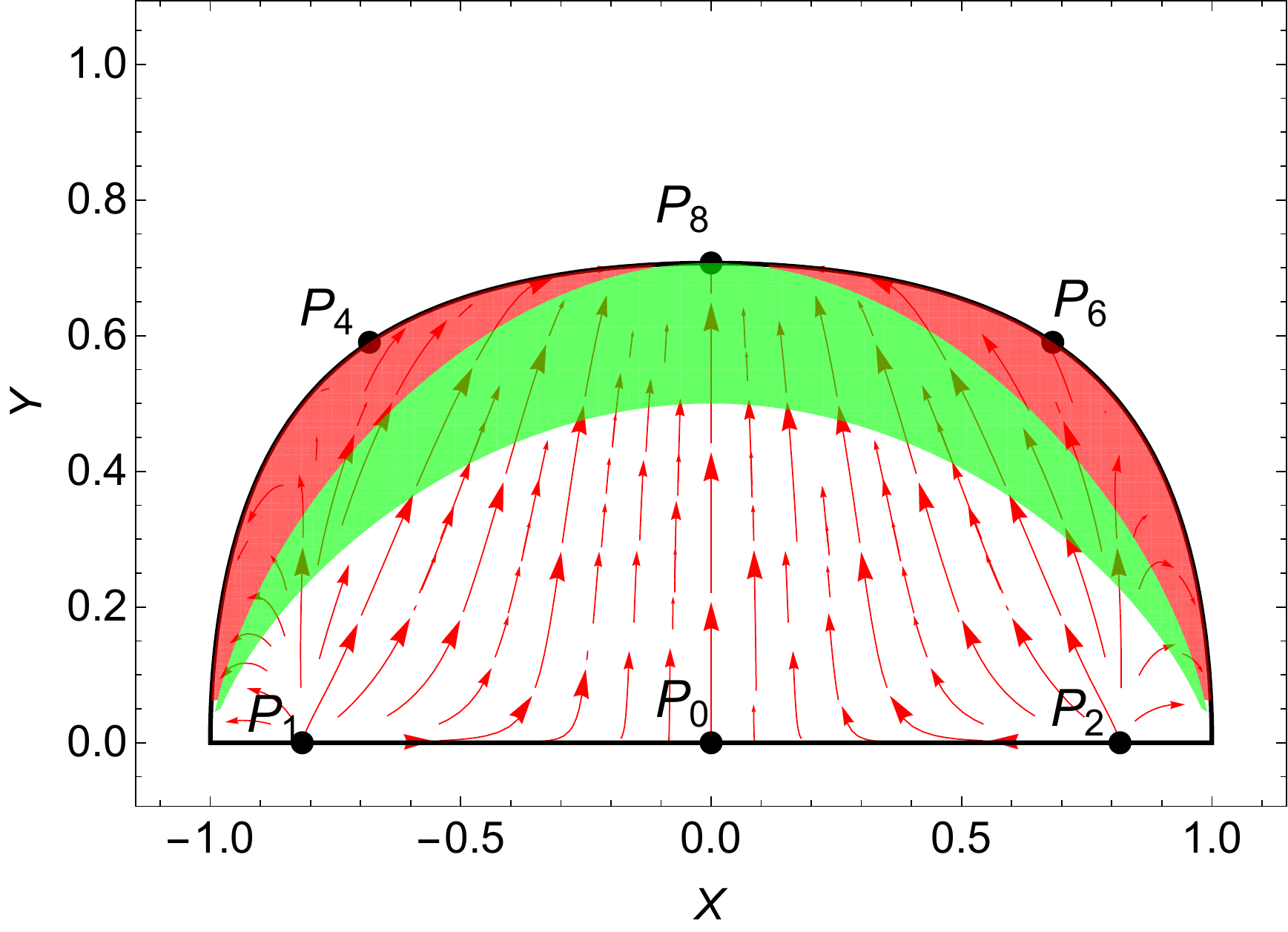}
	\caption{The phase space plot for $\epsilon= -1, \omega_{0} =-3	$ and $\omega_{1} =1/2$. The phase space is constrained, in the green region we have $ -1 \le \omega_{\rm tot} < -1/3 $ and in the red region $ \omega_{\rm tot}<	 -1 $.}
	\label{fig: 3rd para_poincare_phase}
\end{figure}

In the phase space Fig.\ref{fig: 3rd para_poincare_phase} we have plotted $ X $ vs $ Y $. The phase space is constrained and there exist three separate regions. In this case, $ P_{1,2} $ are the non-accelerating points having an equation of state zero. All the vector fields originating from these points are attracted towards $ P_{4}, P_6, P_8 $. Although the system near these points show phantom nature, none of them are stable fixed points except $P_8$. In this system only the point $ P_{8} $ is a stable attractor point and  all the nearby trajectories are attracted towards it and hence it is a global attractor.

\begin{figure}[H]
	\centering
	\includegraphics[scale=0.5]{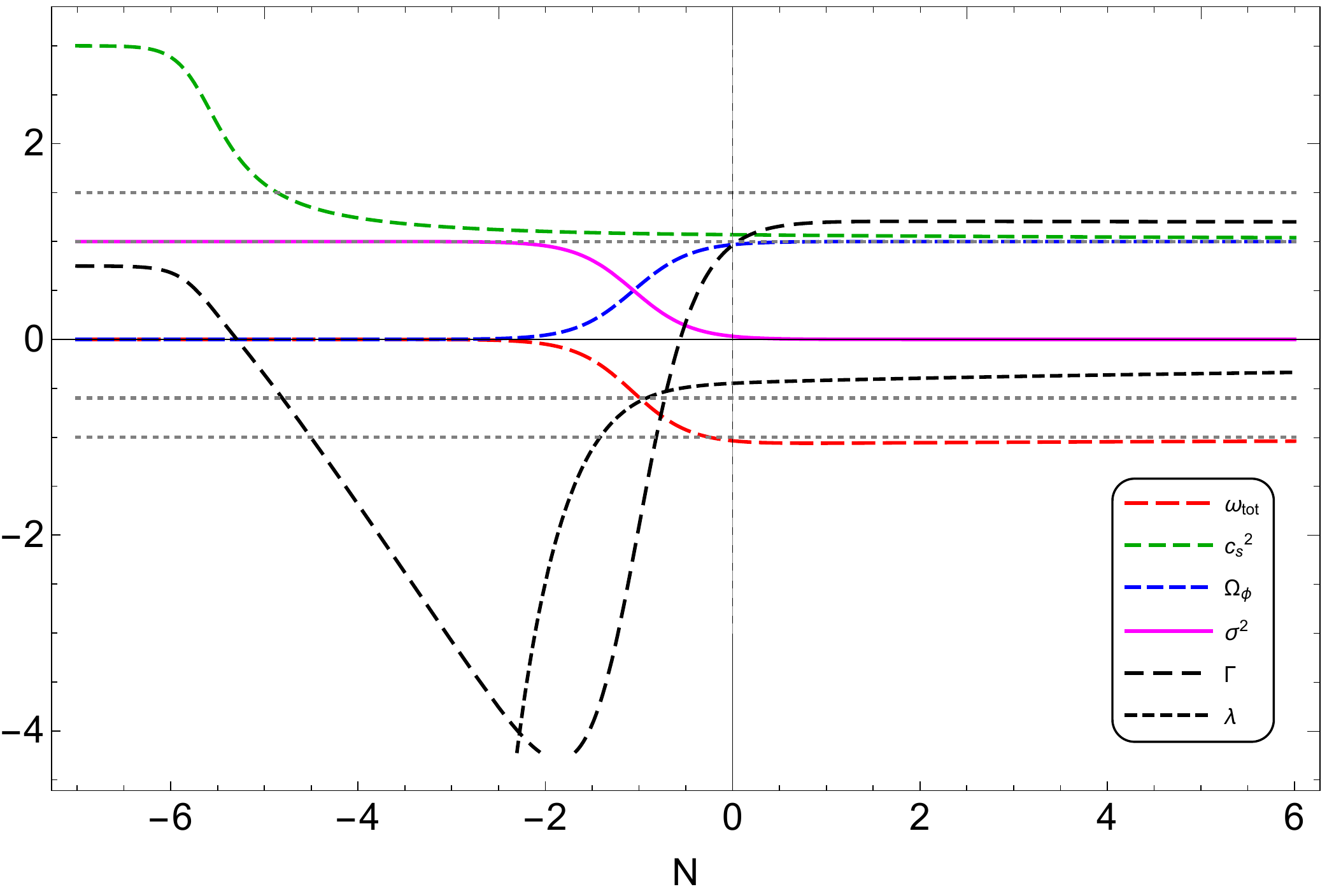}
	\caption{The Evolution plot for $\epsilon= -1, \omega_{0} =-3	$ and $\omega_{1} =1/2$.  }
	\label{fig: Evo_para_2_minus_eps}
\end{figure}

The physical nature of the system can be explained by plotting the dynamical variables against $ N $ as shown in Fig.\ref{fig: Evo_para_2_minus_eps}. We found numerically that at the late time $ P_{8} $ is the stable point and the system evolves towards it. The evolution starts from the distant past with an EoS close to zero. This phase resembles the dark matter dominated regime. In this era the fluid energy density dominates phantom tachyon energy density. Sound speed corresponding to this era is greater than one. As the system evolves the fluid energy density starts decreasing and the tachyon energy density starts to dominate. As a result of this the EoS of the system saturates at $ -1 $ and sound speed becomes $ 1 $. We have also plotted the functions related to the potential: $ \lambda  $ and $ \Gamma $. In the early epoch $\lambda$ decreases exponentially while $\Gamma$ had a controlled behavior. As the system evolves both $ \Gamma $  and $\lambda$ starts increasing. In the transition from dark matter to dark energy phase $\Gamma$ saturates to a value $ 1 $ and $\lambda$ goes to zero. 


\section{Interacting Tachyon with pressureless fluid}

In this section, we extend our analysis to the interacting dark sector scenario i.e. where the tachyon field is coupled with the pressureless matter. Dynamical system analysis of tachyonic field coupled to matter is usually a complicated exercise. We consider a simple interaction of the form
\begin{equation}\label{coupling}
    \mathbb{Q} = Q \rho_{b} \dot{\phi} H,
\end{equation}
where $Q$ is a dimensionless constant. \footnote{This form of coupling is very much similar to \cite{di2020interacting}, where a coupling term $ Q \rho_{\rm de} H  $ was shown to reasonably lower the tension between Plank CMB data and DES measurements.} In our case, the coupled conservation equations become 
\begin{subequations}
 \begin{eqnarray}
&&	\dot{\rho}_{b} + 3 H \rho_{b} = \mathbb{Q} = Q \rho_{b } \dot{\phi} H,\\
&&	\dot{\rho}_{\phi} + 3 H (P_{\phi} + \rho_{\phi}) = - \mathbb{Q} = - Q \rho_{b } \dot{\phi} H.
\end{eqnarray}   
\end{subequations}
The coupling term modifies the field equation \eqref{phi} as,
 \begin{equation}\label{field_coupled}
	\ddot{\phi} + 3 H ( 1- \epsilon \dot{\phi} ^2 ) \dot{\phi} + \frac{\epsilon V_{,\phi}}{V} (1 - \epsilon \dot{\phi}^2) = - \frac{\epsilon Q H \rho_{b}}{V} \left( 1- \epsilon \dot{\phi}^2\right) ^{3/2}
\end{equation}
From the above equation and using the dimensionless variables in Eq.~\eqref{dyn_var}, the dynamical equation for $x$ can be written as
\begin{equation}
	x' = - \epsilon Q \frac{\sigma^2}{y^2}(1 - \epsilon \ x^2 )^{3/2} + \epsilon \lambda \sqrt{3} y (1 - \epsilon \ x^2) - 3x (1 - \epsilon \   x^2),
\end{equation}
Or, using the constraint equation \eqref{constraint_1},
\begin{equation}\label{fied_x}
	x' = - \epsilon Q \left(\frac{(1-\epsilon x^2)^{3/2}}{y^2} - (1-\epsilon x^2)\right) + \epsilon \lambda \sqrt{3} y (1 - \epsilon \ x^2) - 3x (1 - \epsilon \   x^2).
\end{equation}
The dynamical equations of $y$ and $\lambda$ remains the same. 

Note that, in general, a coupling term is expected to give rise to an additional dimensionality in the phase space, requiring us to define an additional dynamical variable. In fact, a generic coupling term may not even allow us to close the system to write it in an autonomous form. Our specific choice of the coupling term in Eq.\eqref{coupling}, however, does not require us to define an additional dynamical variable. The dimensionality of the phase space remains the same. 

We now proceed to investigate the compatibility of our equation of state parametrizations with the choice of this coupling. We note that the issue with the Taylor series parametrizations that we outlined in section \ref{subsec:taylor_param} is independent of whether there is dark sector coupling or not; this is an inherent issue with the parametrization itself while trying to recast it as a constraint over the phase space. Therefore in what follows, we will concentrate only on the second parametrization introduced in section \ref{sec:2nd_param}. Since our goal here is not an extensive analysis of the tachyonic model but to showcase the applicability of the framework we developed, we will concentrate only on the canonical tachyon case ($\epsilon=+1$).

\subsection{Interacting canonical tachyon with the 2nd parametrization}

Proceeding in the same way one can express $\lambda$ as, 
\begin{eqnarray}\label{lambda_int}
    && \lambda(x,y) = \nonumber\\
    &&  \dfrac{ 2 Q \left(y^2 - \sqrt{1-\epsilon x^2}\right) \left(x^4 +(\omega_{0}+1)(x^2 \epsilon -2)\right) }{4\sqrt{3} y^3 (\omega_0 +1) (1-x^2 \epsilon)} \nonumber\\
    && + \dfrac{3 x y^2 \left(x^2 \epsilon  \left(2 (\omega_{0}+1) \sqrt{1-x^2 \epsilon }+y^2 (\omega_{0}+\omega_{1}+1)\right)
-4 (\omega_{0}+1) \sqrt{1-x^2 \epsilon }+x^4 \left(2 \sqrt{1-x^2 \epsilon }-y^2\right)\right)}{4\sqrt{3}\epsilon y^3 (\omega_0 +1) (1-x^2 \epsilon)} \nonumber\\
&& .
 \end{eqnarray}
 The above expression of $\lambda(x,y)$ has a pole of order three at $y\rightarrow0$. After inserting this in \( x'\) or \(y'\), one finds that both the equations still have a pole of order two at $y \rightarrow 0$. This can be regularized by redefining the time variable as
\begin{equation}\label{}
		dN \mapsto  \ y^2   \ dN. 
\end{equation}
With this time redefinition the new autonomous equation for the canonical tachyon field \( \epsilon =+1\) can be written as 
\begin{subequations}\label{coupled1}
	\begin{eqnarray}
		x' & = & \frac{\bigg[ \dfrac{3}{2} \dfrac{x y^4}{\sqrt{1- x^2}} (-\omega_1 + x^2 -1 - \omega_{0} ) - \sqrt{3} \ \lambda(x,y) \ y^3 (x^2 - 1 - \omega_{0})\bigg]}{2 - (x^2 -1 - \omega_{0}) (2/x^2 + 1/(1-x^2)) }, \label{x_prime_coupled1}\\
		y' & = &  \frac{y}{2} \left[ - \ \sqrt[]{3} \lambda(x,y) \  x \  y^3 - 3 \ {\sqrt[]{1 - x^2}} \ \ y^4 + 3y^2  \right], \label{y_prime_coupled1}
	\end{eqnarray}
\end{subequations}
which is now completely regular at $y\rightarrow0$ because of the existence of the $\lambda(x,y)y^3$ term in both the equations.
The 2D autonomous system presents two invariant submanifolds $x=0$ and $y=0$ (noting that $\lambda(0,y)=0$). Finding the critical points of this system in all generality is challenging since the interaction term greatly complicates the system, as clear from the expression of $\lambda(x,y)$ in Eq.\eqref{lambda_int}. The generic critical point structure of the system will depend on three parameters \( (\omega_{0}, \omega_{1}, Q)\). Changes to these variables can have far-reaching effects on the dynamics.  

Nonetheless, we find that there exist critical points $P_0,\,P_1,\,P_2$ corresponding to decelerated matter dominated phases of expansion ($\omega_{\rm tot}=0$) and fixed points $P_{7,8}$ corresponding to accelerated De-Sitter phases of expansion ($\omega_{\rm tot}=-1$); see Tab.\ref{tab:critical_point_coupled_mod}. These are the same points that appeared in Tab.\ref{tab:critical_point_3rd_para} i.e. for the noninteracting canonical tachyon case. Therefore the introduction of an interaction term, at least in the form of Eq.\eqref{tab:critical_point_3rd_para}, does not affect these critical points. The interaction term destroys the other critical points from the non-interacting scenario, while introducing possible others.

As stated, there can be many other critical points depending on the parameter values $(\omega_0,\,\omega_1,\,Q)$ and it is hard to investigate the critical point structure in all generality. However, what our framework does allow us is to find the value of the unspecified model parameter $Q$ in terms of the model parameters $(\omega_0,\,\omega_1)$, whose value we can get from the observations, for a particular cosmological solution to exist. We explain the procedure below. Let us try to look for critical points $(x_*,y_*)$ whose coordinates satisfy the relation 
\begin{equation}\label{h}
 y_*^2 \sqrt{1- \ x_*^2} = - h,
\end{equation}
\( h \) being a constant. Such fixed points lie on the curve $y^2 \sqrt{1- \ x^2}\, +\, h = 0$ in the $x$-$y$ plane, which is the locus of all the points in the phase space whose cosmology is specified by an equation of state parameter $\omega_{\rm tot}=h$. Next, we adopt the following strategy. We utilize the condition \eqref{h} to replace the combination $y^2 \sqrt{1- \ x^2}$ whenever it arises in Eqs.\eqref{x_prime_coupled1} and \eqref{y_prime_coupled1}. This reduces the complexity of the expressions significantly. Next, we solve algebraically for the critical points from these simplified equations. We will find the coordinates of the critical points in terms of \( (h, \ \omega_{0}, \ \omega_{1},\  Q)\). However, for consistency, we must substitute these coordinates back in Eq.~\eqref{h}. This gives us a condition on the model parameters \( (\omega_{0}, \ \omega_{1},\  Q)\).\footnote{An alternative way to get this condition is to replace $y_*^2$ by $-\frac{h}{\sqrt{1- \\ x_*^2}}$ and solve the equation $x'\vert_{x_*}=0$ to obtain $x_*=x_*(\omega_0,\omega_1,Q)$, and then put $x_*=x_*(\omega_0,\omega_1,Q),\,y_*^2 = -\frac{h}{\sqrt{1- \ x_*^2}}$ in the equation $y'\vert_{(x_*,y_*)}=0$.} Only when the model parameters satisfy this particular consistency condition, a critical point satisfying the condition \eqref{h} can exist. In our framework, we use the numerical values of $(\omega_0,\,\omega_1)$ obtained from the observations, whereas $Q$ is still unspecified. Therefore, with this approach, given a pair of values for $(\omega_0,\,\omega_1)$ one can determine the value of $Q$ such that there exists a critical point whose cosmology is specified by $\omega_{\rm tot}=h$. In particular, it is worthwhile to check that, for a given $(\omega_0,\,\omega_1)$, for what values of $Q$ can our interaction term in Eq.\eqref{coupling} allow for the existence of other possible matter-dominated phases ($h=0$) and other possible accelerated phases ($-1\leq h<-1/3$).

\begin{table}[t]
	\centering
	\begin{tabular}{cccc}
		\hline 
		\multicolumn{3}{c}{Critical points for  $ \epsilon = +1$. }\\
		\hline 
		Points & $ x $ & $ y $ & $\omega_{\rm tot}$ \\
		\hline
		$ P_0 $ & $ 0 $ & $ 0 $ & $ 0 $ \\
		\hline
		$ P_{1,2} $ & $\mp \sqrt{1 + \omega_{0}} $ & $ 0 $ & $ 0 $ \\
		\hline
		$ P_{3} $	& $ -\frac{3 h}{Q} $ & $\left[\frac{2Q}{3}\left(\frac{1+h}{-h}\right)\left(\frac{9h^2 - Q^2 (\omega_{0}+1)}{9h^2 - Q^2 (\omega_{0}+\omega_{1}+1)}\right)\right]^{1/4}$ & $h$ \\
		\hline
		$ P_{7,8} $ & $ 0 $ & $ \mp1 $ & $ 0 $\\
		\hline
	\end{tabular}
	\caption{2nd parametrization critical points. }
	\label{tab:critical_point_coupled_mod}
\end{table}
Following the above strategy we find another critical point $P_3$ (see Tab.[\ref{tab:critical_point_coupled_mod}]). We remind the reader that this critical point $P_3$ is \emph{not} the same $P_3$ that appeared in Tab.\ref{tab:critical_point_3rd_para}. The consistency condition between the model parameters for the existence critical point $ P_{3} $ is obtained by putting its coordinates back in Eq.~\eqref{h}, and it can be expressed as 
\begin{equation}\label{consistency}
    (2+h)\left(\frac{9h^2}{Q^2}\right)^2 + \left[(1+\omega_{0}+\omega_{1})h - 2(1+h)(2+\omega_{0})\right]\left(\frac{9h^2}{Q^2}\right) + 2(1+h)(1+\omega_{0}) = 0.
\end{equation}
Within the observationally allowed range of model parameters, $1+\omega_0 > 0$. Also, let us confine ourselves to the case when the critical point $P_3$ is not phantom i.e. $h>-1$. Then, two real positive roots for $9h^{2}/Q^2$ exist when the following condition is met 
\begin{equation}\label{cond}
(1+\omega_{0}+\omega_{1})h - 2(1+h)(2+\omega_{0}) \leq -\sqrt{8(1+h)(2+h)(1+\omega_0)}.    
\end{equation}
Henceforth we focus on the case $ Q>0 $; one can follow similar steps for $ Q<0$. Provided that the condition \eqref{cond} is satisfied, for any given $\omega_0$ and $\omega_1$, one gets two possible values of $Q$ so that the critical point $P_3$ can exist: 
\begin{eqnarray}\label{Qmp}
	\tiny
	Q_{\mp} &= &\frac{3}{2 \sqrt{(h+1) (\omega_{0}+1)}} \bigg[\left( h^3 (\omega_{0}-\omega_{1}+3)+2 h^2 (\omega_{0}+2)\right) \mp \\ \nonumber
	 && 	 \sqrt{h^4 \left(h^2 \omega_{1}^2-2 h \omega_{1} ((h+2) \omega_{0}+3 h+4)+(h (\omega_{0}-1)+2 \omega_{0})^2\right)}  \bigg] ^{1/2}
\end{eqnarray}
Based on \( Q_{-}, Q_{+}\), one actually gets two different versions of $P_3$, which we denote by \( P_{3-}, P_{3+}\) respectively. 

\begin{figure}[H]
 	\begin{minipage}[b]{0.5\linewidth}
 		\centering
 		\includegraphics[scale=0.9]{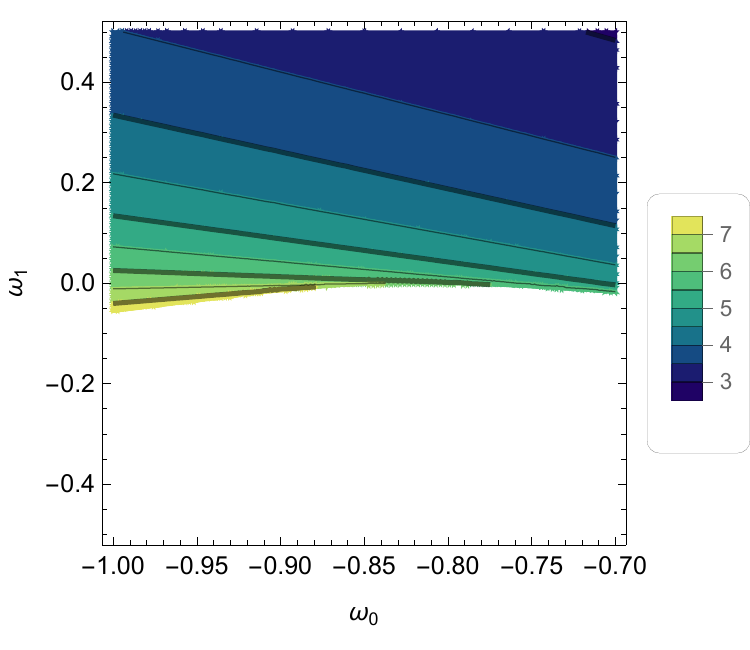}
 		\subcaption{The constraint on \( Q_{-}\) for \( h= -0.9\)}
 	\end{minipage}
 \hspace{0.1cm}
 \begin{minipage}[b]{0.5\linewidth}
 	\centering
 	\includegraphics[scale=0.9]{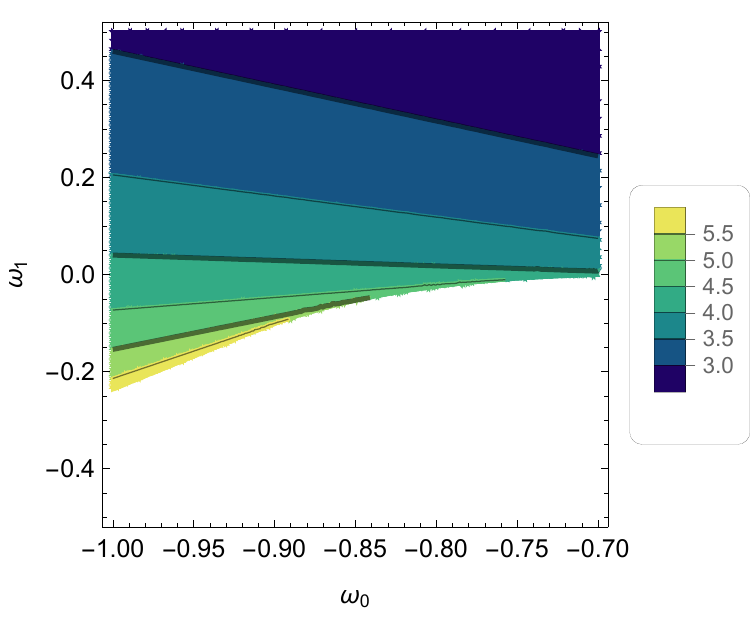}
 	\subcaption{The constraint on \( Q_{-}\) for \( h= -0.8\)}
 \end{minipage}
\vspace{0.05cm}
\begin{minipage}[b]{0.5\linewidth}
	\centering
	\includegraphics[scale=0.9]{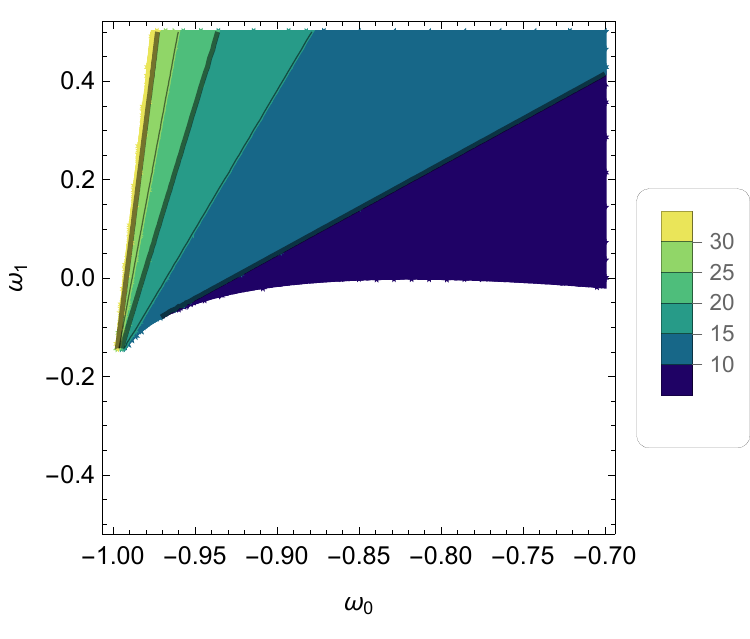}
	\subcaption{The constraint on \( Q_{+}\) for \( h= -0.9\)}
\end{minipage}
\hspace{0.0cm}
\begin{minipage}[b]{0.5\linewidth}
	\centering
	\includegraphics[scale=0.9]{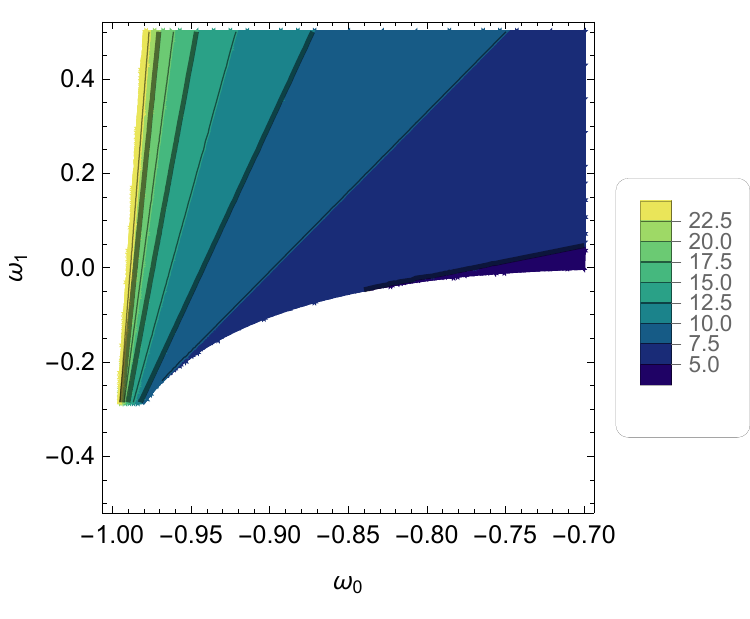}
	\subcaption{The constraint on \( Q_{+}\) for \( h= -0.8\)}
\end{minipage}
\caption{The color pallets show the range of values of the interaction parameters \( Q_{-}, Q_{+}\) against the model parameters $\omega_{0}$ and $\omega_{1}$ such that the critical point $P_3$ may represent accelerated expansion phases with $\omega_{\rm tot}=h=(-0.8,-0.9)$. The entire shaded region is the region specified by the constraint \eqref{cond}.}
\label{fig:constraint_Q}
 \end{figure}

\begin{figure}[H]
	\begin{minipage}[b]{0.5\linewidth}
		\centering
		\includegraphics[scale=0.7]{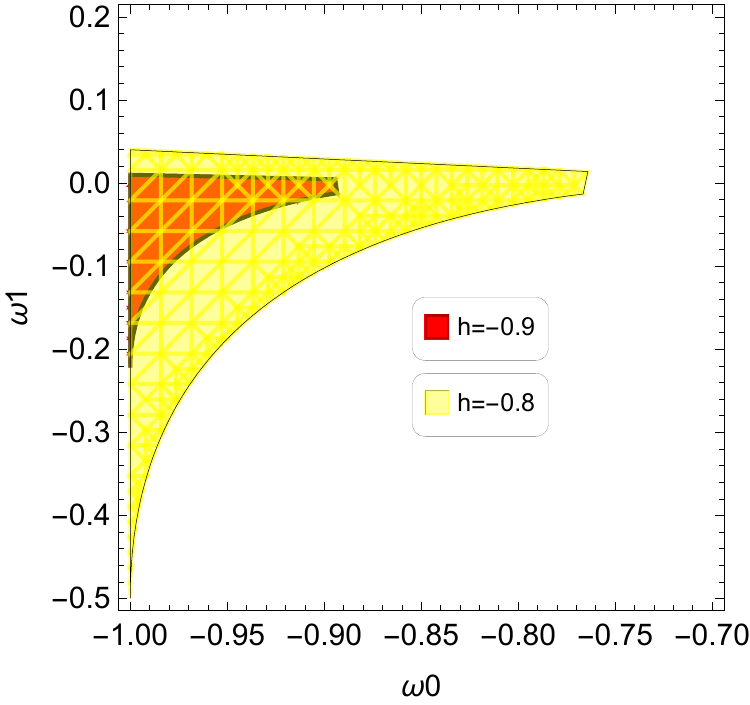}
		\subcaption{The existence of point \( P_{3-}\) for \( h= -0.9,-0.8\).}
	\end{minipage}
	\hspace{1cm}
	\begin{minipage}[b]{0.5\linewidth}
		\centering
		\includegraphics[scale=0.7]{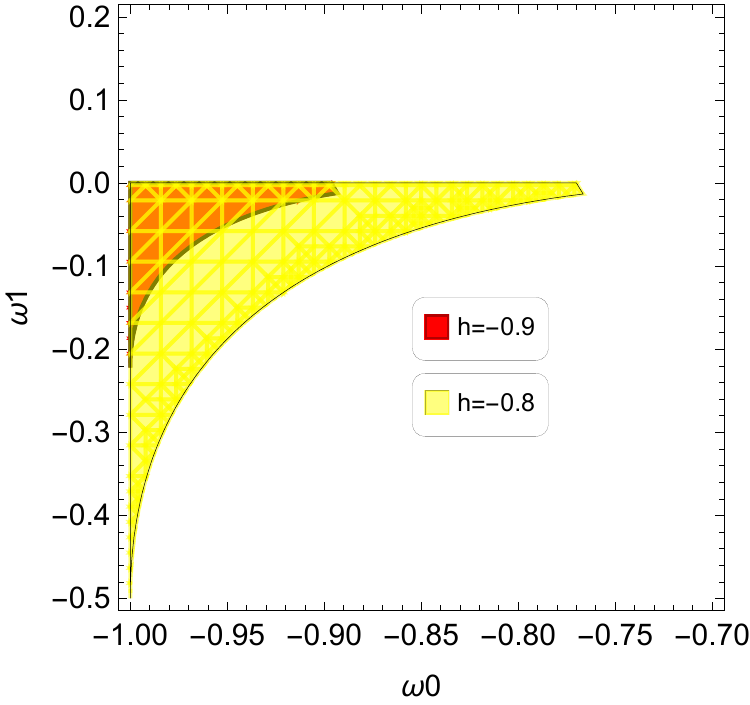}
		\subcaption{The existence of point \( P_{3+}\) for \( h= -0.9,-0.8\).}
	\end{minipage}
\caption{The range in the parameter space $\omega_{0}-\omega_{1}$ such that the critical points $ P_{3-} , P_{3+} $ can exist.}
\label{fig:existence_of_cric_2}
\end{figure}

In Fig.[\ref{fig:constraint_Q}], we show using color pallets the range of values of the interaction parameters \( Q_{-}, Q_{+}\), against the model parameters $\omega_{0}$ and $\omega_{1}$ such that the critical point $P_3$ may represent accelerated expansion phases with $\omega_{\rm tot}=h=(-0.8,-0.9)$.  We emphasize that the condition \eqref{cond}, which is the condition for having real values of $Q_\mp$, by itself is only a \emph{necessary, but not sufficient} condition for critical points $P_{3\mp}$ to exist. One needs to put the values of $Q_\mp$ from Eq.\eqref{Qmp} back in the coordinates of $P_3$ in Table.\ref{tab:critical_point_coupled_mod} and demand that the $y$-coordinate is real. This gives a further constraint on the allowed values of the model parameters $(\omega_{0},\,\omega_{1})$ for critical points $ P_{3\mp} $ to exist, which we show in Fig.[\ref{fig:existence_of_cric_2}]. This is the region in the parameter space $(\omega_0,\,\omega_1)$ such that it is possible for the interaction model of Eq.\eqref{coupling} to support, for some value of the interaction parameter $Q$, the existence of critical points $P_{3-}$ or $P_{3+}$ with $\omega_{\rm tot}=h$. 
  

  \begin{table}[t]
	\centering
	\begin{tabular}{cccccccc}
		\hline 
		\multicolumn{8}{c}{Critical points for  $ \epsilon = +1$. }\\
		\hline 
		Points & $ x $ & $ y $& $\Omega_\phi$ &  $\sigma ^2$ &  $ \omega_{\rm tot} $ & $ c_{s}^2 $ & Stability\\
		\hline
			$ P_{0} $ & 0 & 0&0& 1& 0& 1 & Unstable\\
		\hline
		$ P_{2} $ &  $ 0.32 $ & 0 & 0 &1 & 0 & $ 0.90 $ & Stable\\
		\hline
		$ P_{3} $ & $ 0.45 $& $ 0.95 $ & $ 1 $ & 0 & $ -0.80 $ & $ 0.80 $ & Stable\\
		\hline
		$ P_{4} $ & $ 0.32 $ & $ 0.68 $& $ 0.71 $ & $ 0.52 $ & $ -0.43 $ & $ 0.89 $ & Saddle\\
		\hline 
			$ P_{8} $ & $ 0 $ & $ 1 $ & 1 & 0& $ -1 $ & $ 1 $ & Saddle\\
		\hline
	\end{tabular}
	\caption{The critical points for \( \omega_{0} = -0.9, \omega_{1} = -0.1, Q = 4, \epsilon =+1\) corresponds to the second parametrization.}
	\label{tab:critical_point_coupled_real}
\end{table}

\begin{figure}[H]
	\centering
	\includegraphics[scale=0.9]{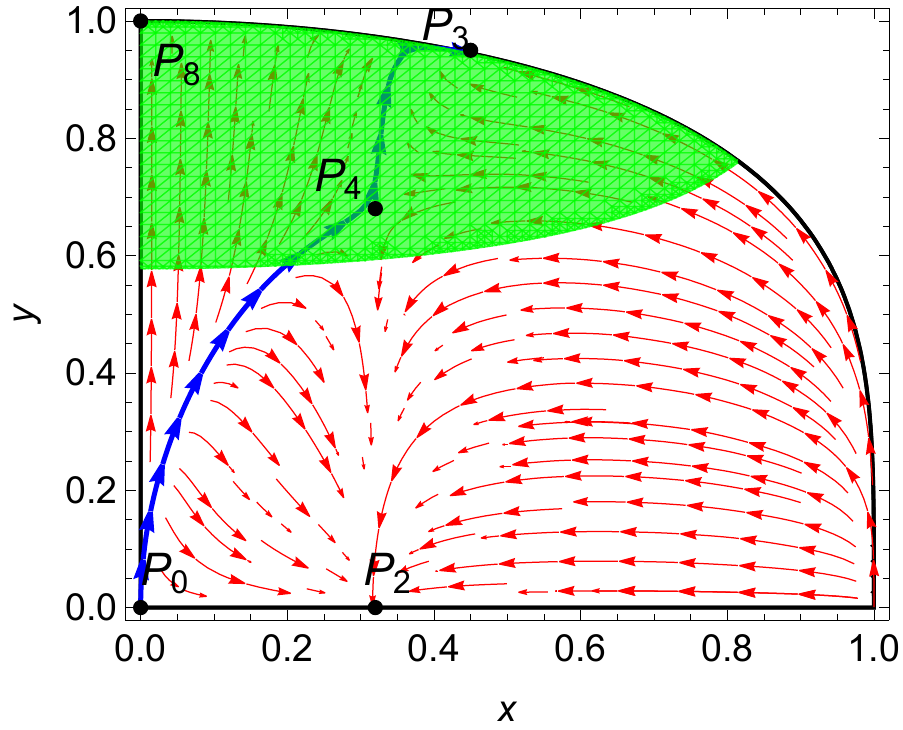}
	\caption{The phase space of the coupled system for \( \omega_{0} = -0.9, \omega_{1} = -0.1, Q = 4, \epsilon =+1\), corresponds to the 2nd parametrization.}
	\label{fig:phase_space_coupled_norm}
\end{figure}

The above strategy, of course, does not allow us to find the entire set of critical points. Given the numerical values of the model parameters $(\omega_0,\,\omega_1,\,Q)$, we can find the critical points numerically. As an example, in Tab.[\ref{tab:critical_point_coupled_real}] we list the physically viable (i.e. allowed by the Friedmann constraint) critical points numerically obtained for the parameter choice \( \omega_{0} = -0.9,\, \omega_{1} = -0.1,\, Q=4\). The corresponding phase space is depicted in Fig.[\ref{fig:phase_space_coupled_norm}], in which the green region shows an accelerating phase and the sound speed is positive and subluminal in the entire region. The points $ P_{0},\,P_{2} $ both represent matter-dominated cosmological epochs. However, $ P_{2} $ is a stable point, so it cannot really represent the actual matter-dominated epoch that our universe has gone through. On the other hand, $ P_{0} $ is a unstable and therefore it can represent a cosmologically relevant matter-dominated epoch. The three points $ P_{3},\,P_{4},\,P_8 $ represent accelerated cosmological epochs, with $P_8$ being a de-Sitter phase. The saddle fixed point $ P_{4} $ is a perfectly viable candidate to characterize the present accelerated epoch of the universe, since the fractional field density is \(\approx 0.71\) and the fractional fluid density is \( \approx 0.29\). Fig.[\ref{fig:phase_space_coupled_norm}] shows the existence of heteroclinic trajectories connecting the matter-dominated unstable phase $ P_{0} $ to the stable accelerated phase $ P_{3} $ through the present intermediate phase $ P_{4} $, representing a possible viable course of evolution for our universe. The stable point $ P_{3} $, which is an accelerated phase dominated completely by the tachyonic field ($\Omega_{\phi}=1$), represents the future asymptotic of the cosmological evolution. 

\begin{figure}[H]
	\centering
	\includegraphics[scale=0.6]{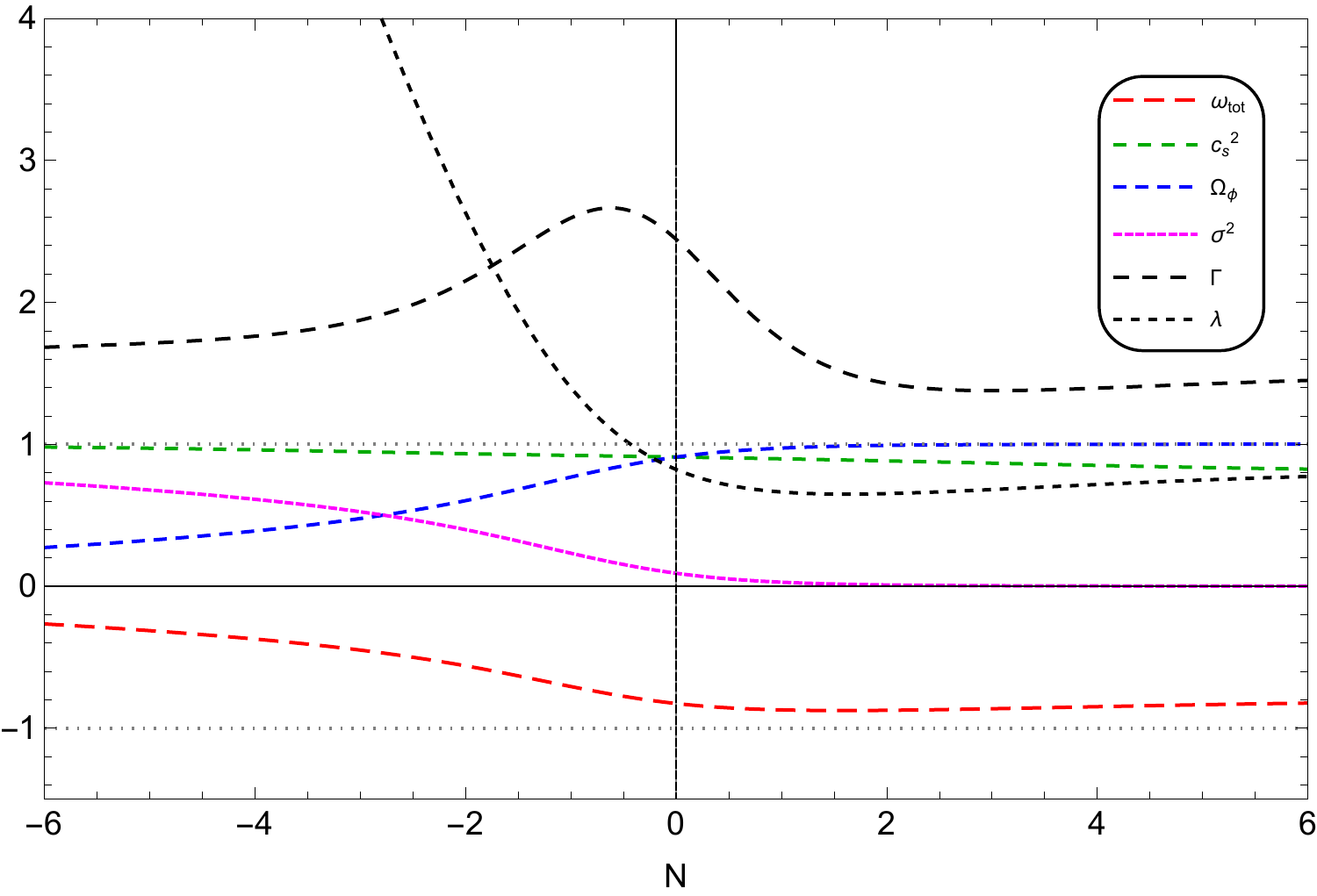}
	\caption{The evolution of the coupled system for \( \omega_{0} = -0.9, \omega_{1} = -0.1, Q = 4, \epsilon =+1\), corresponds to the 2nd parametrization. }
	\label{fig:evo_coupled_norm}
\end{figure}

We investigated the dynamics of the coupled system depicted in Fig.[\ref{fig:evo_coupled_norm}] for some initial conditions. In the early phase, the fluid density outweighs the field density. The total Eos is close to zero, which may mimic the matter phase. In the minimally coupled field-fluid system, the total Eos remains zero at this phase; however, in coupled system, the field density does not dilute to zero. Consequently, the Eos is not zero.  In this phase, $\Gamma$, sound speed $ c_{s}^2 $ and $\lambda$ have non-zero values. As the field density grows, $\Gamma$ demonstrates a rising trend before saturation at some finite value in the late time. Throughout this phase transition, $\lambda$ has a declining trend and becomes saturated.  In the late-time phase, the field density dominates over the fluid density. As a result, the total Eos of the system approaches \(-0.8\). The sound speed remains close to \(1\).


\section{Conclusion}

In this paper we have studied the dynamics of tachyon dark energy models for both the canonical and phantom scalar field. Tachyon dark energy models were studied for more than decades but here we have studied it from a completely new perspective. In all of the previous studies the tachyon potential was assumed to be some function of the tachyon field or the potential parameter $\Gamma$ was assumed to have some preassigned value. Once these information was known, the dynamics and EoS of the evolving system was calculated. In the present paper we have not assumed any prior form of the tachyon potential, rather we have considered parametrization of the EoS of the scalar field. Our work is motivated by various approximate forms of the EoS of the tachyon field. 

It is seen that once we introduce an approximate EoS for the tachyon field the autonomous equations guiding the system changes its form. One of the dynamical variables of the system, namely $\lambda$, becomes redundant, as we do not require to explicitly solve for $\lambda$ to predict the dynamics of the system. Consideration of the parametrization of EoS of scalar field reduces the phase space dimension of the system by one and ultimately a 3D system reduces to a 2D one. This dimensional reduction of the system on the other hand gives a functional form of $\Gamma$ which in general can include a large class of potentials. We studied important dynamical characteristics of the evolving system using our prescription together with the evolution of the important cosmological parameters. The method applied is self-consistent and consequently we can smoothly use the phenomenologically motivated approximate EoS of the tachyon field. Our method is fairly general and we expect that this method can be applied in other dynamic dark energy models.

To show the effectiveness of our proposal we have used two kinds of parametrization to write the EoS of the tachyon field. The first parametrization is more like a Taylor series expansion of the EoS around the present time. In this Taylor series we only retain the first two terms and use it to solve the dynamical system. In solving the autonomous equations, in some of the cases, we had to redefine time and use compactified variables to figure out the critical points of the system. The cosmological dynamics of corresponding to this parametrization shows certain limitations. It is seen that one of the phase space variables, $x(N)$, is either defined only in the future or it is defined only in the past. This property, of the Taylor series parametrization, is a serious limitation of this particular form of the approximate EoS.

To obtain a viable cosmology we have used another form of parametrization of the EoS of the tachyon field. In the second parametrization the EoS of the tachyon field is expressed as a sum of a constant term and a time dependent part. The EoS  closely resembles dark energy EoS when the Hubble parameter tends to be a constant. We have presented the solutions in a detailed manner showing the dynamical evolution of various cosmological variables. In the case of phantom tachyons it is seen that the sound speed in the scalar field sector persistently remains superluminal. We have also verified that superluminal sound propagation in this cases cannot be avoided. This kind of behavior arises in some k-essence models of dark energy. In principle this superluminal speed does not break Lorentz invariance as this speed is only appropriately observed in a particular tachyonic background. 

Although a particular form of the potential of the tachyon field was not assumed apriori or nor have we assumed any particular functional form of $\Gamma$ to start with, we have described an approximate method using which one can figure out the form of the tachyon potential $V(\phi)$. The method is approximate as because the the form of the potential may not match the exact potential form at all cosmological phases. In the example we have shown two approximate forms of the potential which matches with the desired exact potential in the late phase of cosmic evolution.  In principle one can use more accurate techniques to find out the functional form of the potential but as the potential does not play any primary role in our analysis we have not attempted to do so. 

The applicability of our approach is not confined to non-interacting scenarios only. We have shown it explicitly by considering a simple example of tachyon and dark matter interaction. In general, the interaction terms call for the introduction of an additional dynamical variable, increasing the dimensionality of the phase space. However, for the specific model we considered here \eqref{coupling}, namely $\mathbb{Q} = Q \rho_{b} \dot{\phi} H$, we did not have to introduce an additional dynamical variable. The dimensionality of the phase space remains the same. Given the numerical values of the model parameters $(\omega_0,\,\omega_1)$, our framework allows us to determine the interaction parameter $Q$ so that the interacting model given by Eq.\eqref{coupling} can support the existence of specific cosmological epochs. For given numerical values of the parameters $(\omega_0,\,\omega_1,\,Q)$, one can analyze the dynamics of the phase space without explicitly specifying the potential of the tachyon field. In principle, our framework can also be extended to incorporate more complicated interacting models.

In summary, we have presented a new perspective to the dynamical system analysis of tachyon dark energy model. Our approach is generic and can be applied to other scalar field dark energy models.

\section{Acknowledgement}

SC acknowledges funding support from the NSRF via the Program Management Unit for Human Resources and Institutional Development, Research and Innovation [grant number B01F650006].

\vskip .2cm
\newpage
\appendix
\centerline{\bf \Large Appendix}

\section{Calculation of \texorpdfstring{$\Gamma$}{Lg} for the first parametrization}
\label{appen.1}

Taking the derivative of above equation with respect to the $ N $ and using Eq.~\eqref{lam_prime}, we will get $ \Gamma  $ in terms of $ x $ and $ y $ as 
\begin{multline}\label{}
	\Gamma = \frac{1}{\epsilon  \left(-6 x^5 \epsilon ^2+6 x^3 \epsilon +x \omega_{1}\right)^2} \bigg[ 36 x^{10} \epsilon ^5-18 x^8 \epsilon ^4 \left(y^2 \sqrt{1-x^2 \epsilon }+3\right) + 18 x^6 \epsilon ^3 \left(2 y^2 \sqrt{1-x^2 \epsilon }-\omega_{1}\right) \\
	-x^2 \omega_{1} \epsilon  \left(3 y^2 \sqrt{1-x^2 \epsilon }+2 \omega_{1}+3\right)+3 x^4 \epsilon ^2 \left(y^2 (\omega_{1}-6) \sqrt{1-x^2 \epsilon }+7 \omega_{1}+6\right)+\omega_{1}^2
	\bigg]
\end{multline}
For the phantom case after the compactification the $\Gamma$ can be expressed as, 
\begin{multline}\label{}
	\Gamma = \dfrac{1}{\left(Y^2-1\right) \left(X^5 \omega_{1}-2 X^3 (\omega_{1}+3)+X \omega_{1}\right)^2} \bigg[-18 X^4 \sqrt{1-X^2} Y^2+18 X^4 \left(3 X^2-1\right) \left(Y^2-1\right)\\
	+3 X^2 \sqrt{1-X^2} Y^2 \omega_{1}-6 X^4 \sqrt{1-X^2} Y^2 \omega_{1}+ 3 X^6 \sqrt{1-X^2} Y^2 \omega_{1}-3 \left(5 X^2+1\right) \\
	\left(X-X^3\right)^2 \left(Y^2-1\right) \omega_{1} -\left(X^2-1\right)^4 \left(X^2+1\right) \left(Y^2-1\right) \omega_{1}^2
	\bigg]
\end{multline}
From here we found that $ \Gamma = 1 $ at critical points $ (X,Y) = (1,0) $ and becomes infinite for the other critical point. 


\section{Stability of the fixed points for the case of phantom tachyons in the first parametrization}
\label{app:stability}

The fixed points listed in Tab.\ref{tab:epslm1} are non-hyperbolic. Therefore their stability cannot be determined via the usual Jacobian analysis. In these situations one can resort to a more formal center manifold analysis but we notice that there is a way around here. The trick is to look for invariant submanifolds of the system described by Eq.\eqref{ds_param1_eps-1}. From Eq.\eqref{ds_param1_eps-1} we can directly observe that $X=1$ and $Y=0$ are invariant submanifolds of the system, because
\begin{equation}
    X'\vert_{X=1}=0, \qquad Y'\vert_{Y=0}=0.
\end{equation}
There is another invariant submanifold of the system which is not so clearly identified from Eq.\eqref{ds_param1_eps-1}, namely $\sigma=0$. To see that this is indeed an invariant submanifold one can start from the definition of $\sigma$ in Eq.\eqref{dyn_var} and take it's derivative with respect to $N=\ln(a)$. Using the time redefinition in Eq.\eqref{time_redef_2} and the definition of the compact dynamical variables in Eq.\eqref{dyn_var_comp} and after some straightforward calculations we arrive at:
\begin{equation}\label{dyn_sigma}
    \sigma' = - \frac{3}{2}\sigma X Y^2 \frac{\sqrt{1-X^2}}{1-Y^2},
\end{equation}
where $'$ is to be understood as $\frac{d}{d\bar{N}}$. Clearly $\sigma=0$ is an invariant submanifold as
\begin{equation}
    \sigma'\vert_{\sigma=0}=0.
\end{equation}
In fact one could have already guessed the existence of this invariant submanifold from the physical argument that, if the universe starts as a vacuum it remains so. There is no mechanism in classical gravity by which matter can be produced out of vacuum.

At the vicinity of $X=1$, $X'>0$ ($X'<0$) for $\omega_{1}<0$ ($\omega_{1}>0$). Correspondingly the invariant submanifold $X=1$ is attracting nearby phase flows towards it for $\omega_1<0$ and repelling nearby phase flows away from it for $\omega_1>0$. At the vicinity of $Y=0$ we have, to the leading order:
\begin{equation}
    Y'\approx \frac{1}{2}XY\left(3-\frac{\omega_1}{2}(1-X^2)^2\right).
\end{equation}
In particular, at the vicinity of the fixed point $P_1\equiv(X,Y)=(1,0)$, $Y'\approx\frac{3}{2}XY$, which is always positive in the first quadrant. Therefore in the vicinity of $P_1$ the invariant submanifold $Y=0$ is always repelling nearby flows away from it. At the vicinity of $\sigma=0$, the flow is always towards $\sigma=0$, since the coefficient of $\sigma$ on the right hand side of Eq.\eqref{dyn_sigma} is always negative in the first quadrant. Therefore the curve given by Eq.\eqref{curve_2}, which represents $\sigma^{2}=0$, i.e. vacuum cosmologies, is attracting nearby flows.

The fixed point $P_1$ lies at the intersection of the invariant submanifolds $X=1$ and $Y=0$. Therefore this fixed point is a saddle for $\omega_1<0$ and a repeller for $\omega_1>0$. The fixed point $P_2$ lies at the intersection of the invariant submanifolds $X=1$ and $\sigma=0$. Therefore this fixed point is an attractor for $\omega_1<0$ and a saddle for $\omega_1>0$.


\section{Calculation of \texorpdfstring{$\Gamma$}{Lg} for the second parametrization}
\label{app:norm_tachyon_2nd_para}

In this appendix we present the expressions of $\Gamma$ for the various values of $\epsilon$ in the second parametrization of the EoS for the scalar field.
\subsection{Analysis of \texorpdfstring{$\lambda$}{Lg}  and \texorpdfstring{$\Gamma$}{Lg} for normal Tachyon (\texorpdfstring{$\epsilon = 1$}{Lg} ) case. }

The expression of $ \Gamma $ in this case is: 
\begin{multline}\label{Gamma}
	\Gamma = \dfrac{1}{\left(-4 \sqrt{1-x^2} x (\omega_{0}+1)+x^5 \left(2 \sqrt{1-x^2}-y^2\right)+x^3 \left(2 \sqrt{1-x^2} (\omega_{0}+1)+y^2 (\omega_{0}+\omega_{1}+1)\right)\right)^2} \times \\
	-16 x^{12}+4 x^{10} \left(-4 \sqrt{1-x^2} y^2+y^4+9\right)-8 (\omega_{0}+1)^2 \left(\sqrt{1-x^2} y^2-1\right)+\\
	2 x^2 (\omega_{0}+1) \left(\sqrt{1-x^2} y^2 (13 \omega_{0}+3 \omega_{1}+13)+y^4 (-(\omega_{0}+\omega_{1}+1))-6 (\omega_{0}+1)\right)+\\
	x^8 \left(4 \sqrt{1-x^2} y^2 (3 \omega_{0}+3 \omega_{1}+8)+y^4 (-(4 \omega_{0}+6 \omega_{1}+9))-4 (\omega_{0}+6)\right)-\\
	x^4 \bigg(2 \sqrt{1-x^2} y^2 (\omega_{0}+1) (13 \omega_{0}+5 \omega_{1}+14)+y^4 \left(-\left(\omega_{0}^2-2 \omega_{0} (\omega_{1}-2)-3 \omega_{1}^2-2 \omega_{1}+3\right)\right)\\
	+4 \omega_{0} (\omega_{0}+1)\bigg)+\\
	2 x^6 \left(\sqrt{1-x^2} y^2 \left(6 \omega_{0}^2+\omega_{0} (4 \omega_{1}+3)-4 \omega_{1}-3\right)+y^4 \left(\omega_{0} (\omega_{1}+2)+\omega_{1}^2+5 \omega_{1}+2\right)+4 (\omega_{0}+1)^2\right)\,.
\end{multline}
Therefore $\Gamma$ solely depends on the dynamical variables $ x $ and $ y $. We have seen that $\lambda$ diverges at $ \omega_{0} \to -1 $ and $ x  \to \pm 1$, while $\Gamma$ becomes finite. Hence $\Gamma = -1$ for $ x \to \pm 1 $. The expression of $\Gamma$ when $\omega_{0} \to -1$ is:
%
\begin{eqnarray}
\Gamma &=&\dfrac{1}{\left(x^3 \left(2 \sqrt{1-x^2}-y^2\right)+x y^2 \omega_{1}\right)^2}
\times \left[ -16 x^8+4 x^6 \left(-4 \sqrt{1-x^2} y^2+y^4+9\right)\right.\nonumber\\
  &-&\left.3 y^4 \omega_{1}^2+2 x^2 y^2 \omega_{1} \left(y^2 (\omega_{1}+4)-8 \sqrt{1-x^2}\right)\right.\nonumber\\
&+&\left.x^4 \left(4 \sqrt{1-x^2} y^2 (3 \omega_{1}+5)+y^4 (-(6 \omega_{1}+5))-20\right)\right]\,.
\end{eqnarray}

\subsection{Expression of \texorpdfstring{$\lambda$}{Lg} and \texorpdfstring{$\Gamma$}{Lg} for the phantom tachyon field \label{app:phan_tach_lamb_second}}

In the transformed variable $\lambda$ can be written as: 
\begin{multline}\label{}
	\lambda = \dfrac{1}{4 Y (\omega_{0}+1) \left(\frac{Y^2-1}{X^2+Y^2-1}\right)^{3/2} \left(X^2+Y^2-1\right)^2} \sqrt{3} X \Bigg[ X^4 \left(\frac{Y^2}{X^2+Y^2-1}+2 \sqrt{\frac{Y^2-1}{X^2+Y^2-1}}\right)\\
	 -4 (\omega_{0}+1) \sqrt{\frac{Y^2-1}{X^2+Y^2-1}} \left(X^2+Y^2-1\right)^2 \\
-X^2 \left(-X^2-Y^2+1\right) \left(2 (\omega_{0}+1) \sqrt{\frac{Y^2-1}{X^2+Y^2-1}}-\frac{Y^2 (\omega_{0}+\omega_{1}+1)}{X^2+Y^2-1}\right)\Bigg]\,.
\end{multline}
One can write \(\Gamma\) as: 
\begin{multline}
	\Gamma = \Bigg[\left(\left(X^2+Y^2-1\right)^5 \left(8 (\omega_{0}+1)^2 \left(-\frac{Y^2 \left(\frac{Y^2-1}{X^2+Y^2-1}\right)^{3/2}}{Y^2-1}-1\right) \right.\right. \\
	+\frac{16 X^{12}}{\left(X^2+Y^2-1\right)^6}-\frac{4 X^{10} \left(\frac{Y^2 \left(\frac{4 \left(Y^2-1\right)}{\sqrt{\frac{Y^2-1}{X^2+Y^2-1}}}+Y^2\right)}{\left(X^2+Y^2-1\right)^2}+9\right)}{\left(X^2+Y^2-1\right)^5}\\
	-\frac{2 X^2 (\omega_{0}+1) \left(-\frac{Y^2 \left(\frac{Y^2-1}{X^2+Y^2-1}\right)^{3/2} (13 \omega_{0}+3 \omega_{1}+13)}{Y^2-1}-\frac{Y^4 (\omega_{0}+\omega_{1}+1)}{\left(X^2+Y^2-1\right)^2}-6 (\omega_{0}+1)\right)}{X^2+Y^2-1}\\
	+\frac{X^8 \left(\frac{4 Y^2 \left(\frac{Y^2-1}{X^2+Y^2-1}\right)^{3/2} (3 \omega_{0}+3 \omega_{1}+8)}{Y^2-1}+\frac{Y^4 (4 \omega_{0}+6 \omega_{1}+9)}{\left(X^2+Y^2-1\right)^2}+4 (\omega_{0}+6)\right)}{\left(X^2+Y^2-1\right)^4}\\
	+\frac{X^4 \left(-\frac{2 Y^2 (\omega_{0}+1) \left(\frac{Y^2-1}{X^2+Y^2-1}\right)^{3/2} (13 \omega_{0}+5 \omega_{1}+14)}{Y^2-1}-\frac{Y^4 \left(\omega_{0}^2-2 (\omega_{0}+1) \omega_{1}+4 \omega_{0}-3 \omega_{1}^2+3\right)}{\left(X^2+Y^2-1\right)^2}+4 \omega_{0} (\omega_{0}+1)\right)}{\left(X^2+Y^2-1\right)^2} \\
	- \left.\left.\frac{2 X^6 \left(\frac{Y^2 \left(\frac{Y^2-1}{X^2+Y^2-1}\right)^{3/2} (-\omega_{0} (6 \omega_{0}+4 \omega_{1}+3)+4 \omega_{1}+3)}{Y^2-1}+\frac{Y^4 (\omega_{0} (\omega_{1}+2)+\omega_{1} (\omega_{1}+5)+2)}{\left(X^2+Y^2-1\right)^2}+4 (\omega_{0}+1)^2\right)}{\left(X^2+Y^2-1\right)^3}\right)\right)   \Bigg] \times \\
	\Bigg[\left(X^5 \left(-\frac{Y^2}{X^2+Y^2-1}-2 \sqrt{\frac{Y^2-1}{X^2+Y^2-1}}\right)+4 X (\omega_{0}+1) \sqrt{\frac{Y^2-1}{X^2+Y^2-1}} \left(X^2+Y^2-1\right)^2+\right. \\
	\left.X^3 \left(-X^2-Y^2+1\right) \left(2 (\omega_{0}+1) \sqrt{\frac{Y^2-1}{X^2+Y^2-1}}-\frac{Y^2 (\omega_{0}+\omega_{1}+1)}{X^2+Y^2-1}\right)\right)^2\Bigg]^{-1}\,.
\end{multline}


\bibliographystyle{unsrt}
\bibliography{tachyon}

\end{document}